\documentclass[11pt]{article}

\usepackage[utf8]{inputenc}
\usepackage{tcolorbox}
\usepackage{pifont}
\newcommand{\cmark}{\ding{51}}
\newcommand{\xmark}{\ding{55}}
\usepackage[final]{acl}
 \usepackage{booktabs}
\usepackage{times}
\usepackage{latexsym}

\usepackage[T1]{fontenc}

\usepackage[utf8]{inputenc}
\usepackage{amsmath}
\usepackage{amsfonts}

\usepackage{microtype}

\usepackage{inconsolata}

\usepackage{graphicx}

\usepackage{soul} 

\usepackage{xurl}         
\title{Make Mechanistic Interpretability Auditable: \\ A Call to Develop Guidelines via Continuous Collaborative Reviewing}

\author{
Michael Lan \\
Martian
\And
Narmeen Fatimah Oozeer \\
Martian
\And
Chaithanya Bandi \\
Martian
\And
Philip Quirke \\
Martian
\AND
Austin Meek \\
University of Delaware \\
MATS
\And
Fazl Barez \\
University of Oxford \\
Martian
\And
Amirali Abdullah \\
ThoughtWorks \\
Martian
}

\begin{document}
\maketitle
\begin{abstract}
While mechanistic interpretability (MI) has produced important insights into neural network internals, the field has yet to establish a standardized system to audit experiments. As such, many of its findings remain underutilized in safety-critical applications such as medical AI and autonomous systems, as stakeholders cannot certify their validity. Recent work demonstrates this concretely: two papers found conflicting conclusions for the same behavior, and a third study revealed that both were partially correct but incomparable due to methodological inconsistencies. Without standardized auditing, such ambiguities hinder adoption in high-stakes contexts requiring strong correctness guarantees. We call for the MI community to work towards developing a novel reviewing system that complements peer review via: (1) Continuous reviewing supported by a \emph{Collaborative Reviewing Platform} where meta-science results and discussions (such as critiques, negative results, post-hoc extensions, reproductions, replications, and partial results) that fit outside of papers are organized and discussed, allowing for comments and revisions to be made at any time (2) Generalizing good practices found on this platform into expert-verified guidelines and protocols to improve auditing efficiency, and (3) Source-based auditing systems that track arguments which claims depend on. This position paper encourages constructive debate over the necessity, design and implementation of such a framework, providing early concrete examples to help catalyze these dialogues. Overall, we propose that auditing MI itself is essential for its application in AI safety, industry, and governance. 
\end{abstract}

\section{Introduction}

\begin{figure}[ht]
  \centering
  \includegraphics[width=0.8\columnwidth]{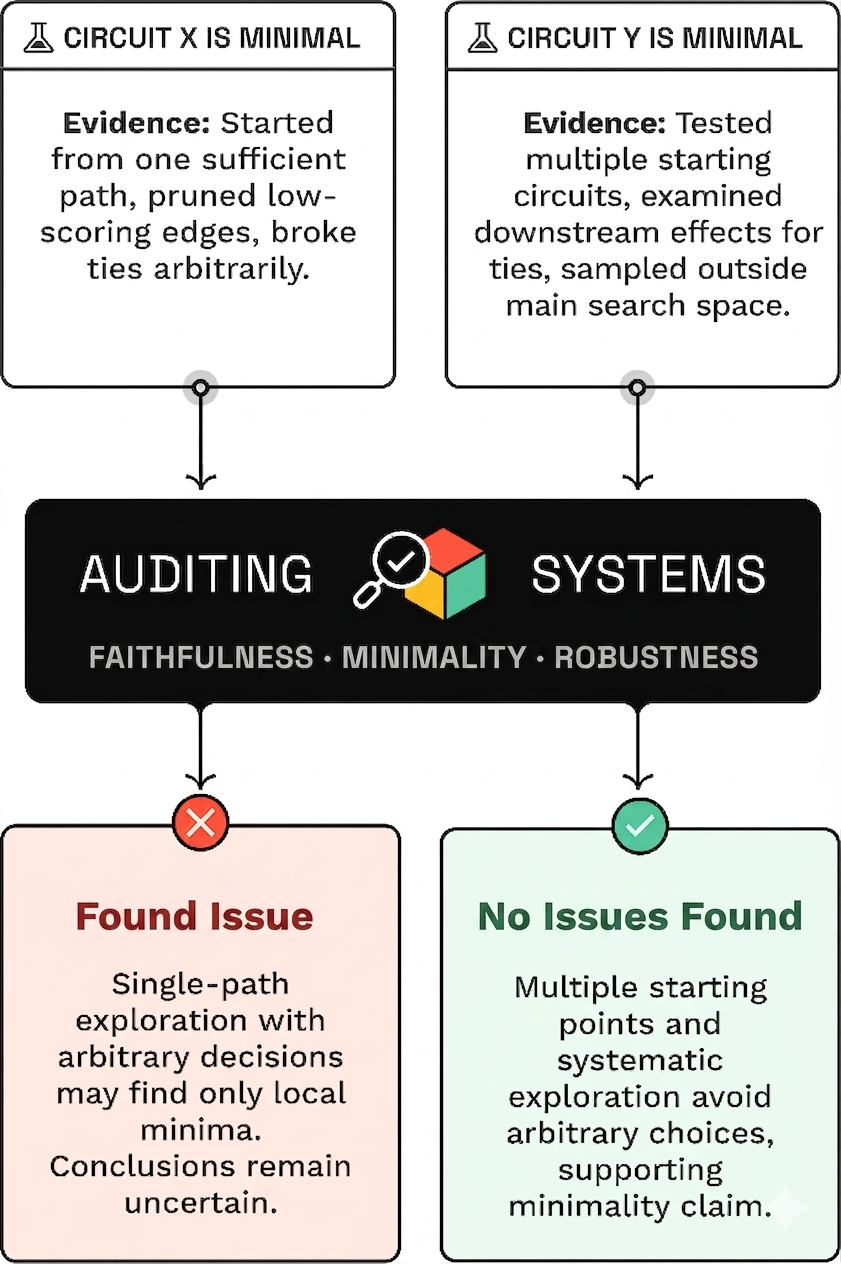}
  \caption{A high-level example of auditing two different hypothesized claims based on community-refined guidelines.}
  \label{fig:image_label}
\end{figure}

Two mechanistic interpretability studies proposed competing explanations for the same mechanism in a neural network \citep{Chughtai2023, Stander2024}. Each was peer reviewed, and both appeared credible. Yet, they reached contradictory conclusions about the internal mechanisms responsible for the model's behavior. Only when a third paper reconciled them under a unifying framework was it revealed that both were partially correct, yet incomparable due to their experimental methodology \citep{wu2025unifiedverifiedunderstandinggroupoperation}. For stakeholders considering deployment in safety-critical applications such as medical diagnostics, autonomous systems, and financial regulation, such ambiguity is unacceptable. Without good auditing protocols, how can decision-makers determine which claims to trust? 

Mechanistic interpretability (MI), the study of discovering algorithmic explanations for the internal workings of neural networks \citep{bereska2024mechanistic}, has advanced rapidly in recent years, producing important insights for practical applications such as model steering \citep{anthropic2025claude}, hallucination detection \citep{obeso2025realtimedetectionhallucinatedentities}, and AI auditing \citep{marks2025auditinglanguagemodelshidden}. These advances have generated significant interest across academia and industry \citep{amodei2025urgency, balsam2024goodfireember}.
However, the field has grown without establishing widely enforced standards for auditing experiments. This risks serious issues: leading MI experts \citep{sharkey2025open} warn that without extensive hypothesis validation, researchers are in danger of conducting studies with significant room for subjective interpretation, lacking objective frameworks to adjudicate competing explanations. This increases the chances of overconfident conclusions and ``interpretability illusions": claims which appear correct, yet fail essential sanity checks \citep{Adebayo2018, Friedman2024}.


These systemic issues have dire consequences for the field, making its findings less likely to be adopted for real-world applications that require strong safety guarantees \citep{wu2025dila, Golgoon_2024}. While studies from leading experts benefit from rigorous internal verification \citep{nanda2023progress, tigges2024llm}, the broader research community lacks this assurance. Without certification procedures, valuable findings remain underutilized, and if these issues are not corrected, the field is at risk of being overwhelmed by studies of uncertain quality.

The MI community has recognized this challenge. \citet{bereska2024mechanistic} emphasizes that objective evaluation requires well-defined metrics, standardized benchmarks, and algorithmic testbeds. 
\citet{Casper2023EIS} argues that MI must adopt engineering rigor for grounding claims. 
\citet{Meloux2025TheDS} argues that AI interpretability methods can produce plausible-looking explanations even for spurious patterns, and therefore reframes interpretability as a statistical inference problem that should quantify uncertainty and test explanations against explicit alternatives.
Recent work argues that MI explanations may be fundamentally non-identifiable, with multiple distinct circuits or abstractions equally matching the same behavior
, underscoring further condideration on what counts as acceptable explanations
\citep{méloux2025everythingeverywhereoncemechanistic}. 
Recent theory also suggests that causal abstraction
can become too permissive to be informative, implying that MI audits must also evaluate assumptions that make an explanation meaningful \citep{sutter2025nonlinearrepresentationdilemmacausal}. \citet{SecondLookResearch} aims to produce open-source replications of MI and AI safety research in a centralized manner, addressing reproducibility issues in research \citep{Semmelrock}.
Yet despite these urgent calls, a widely accepted and consistently enforced approach has yet to be established. 



Due to rapidly evolving research, such that previously declared validity can vastly change in short time periods, we advocate for the research community to pursue the development of \textbf{continuous reviewing} that complements peer review. 
We define continuous reviewing as a collaborative approach that gradually refines pieces of research using \textbf{meta-analysis} results and discussions that fit outside of a paper. 
Meta-analysis includes comments, criticisms, negative results, reproductions, replications, non-novel extensions, minor yet useful results which are not enough for a paper, and partial results. We define these results that complement existing papers but do not exist in paper-form as \textbf{meta-results}. We propose that this development will benefit via improved organization of meta-results and discussions using a \textbf{community-driven platform}.
We posit continuous reviewing as an experimental approach following previous works such as OpenReview \citep{soergel2013open}. 
Useful auditing patterns found on this platform from refining meta-results can eventually transform into standardized empirical guidelines. 

This paper's aim is not to give specific guidelines, but to advocate for the development of an auditing system for MI, arguing this can be done by first improving meta-analysis organization. We propose a concrete implementation roadmap via a community-driven reviewing system built with three important components:
\begin{enumerate} 
  \item \textbf{Continuous Reviewing using Meta-Analysis} supported by a \emph{Collaborative Reviewing Platform} where meta-results and discussions (such as negative results, reproductions, and partial results) that fit outside papers are neatly organized and built upon by researchers, allowing for comments and revisions to be made at any time. 
  \item \textbf{Community-Refined Guidelines and Protocols} generalized from common, useful and expert-supported points developed on this platform that assist in both community reviewing and professional auditing.
  \item \textbf{Source-Based Auditing Systems} that weigh claims by tracing and tracking the assumptions, evidence, specific experiments, and other claims they depend on. 
\end{enumerate}


Crucially, we emphasize that any developed standards should not be seen as the most definitive test of a study's quality; rather, we advocate that they provide one rigorous dimension, out of many, to include in nuanced discussions during study evaluations. This approach helps avoid issues such as a false sense of rigor when a study passes a checklist.
Figure~\ref{fig:image_label} gives an auditing process example.

This position paper seeks to spark constructive discussion about the necessity, design, and implementation of an  auditing framework for MI, introducing concrete examples to help catalyze dialogues, and arguing for continuous community reviewing. By establishing this framework early, we ensure that MI matures into a reliable approach for AI safety, rather than remaining an informal practice.
Our main contributions include:
\begin{itemize}
    \item Arguments establishing the need for MI auditing, using examples to motivate discussions
    \item A plan-to-action for implementing a community-based MI reviewing system, proposing an approach using continuous reviewing via meta-analysis
    \item A proposal for a collaborative platform where researchers propose community-refined guidelines which influence professional auditing guidelines in safety-critical applications
\end{itemize}

\section{Technical Background and Potential MI Pitfalls}
\label{sec:background}

Due to the inherent difficulty of explaining black-box models, MI studies may fall into empirical pitfalls that span across approaches, which can be avoided by following guidelines as described in Table \ref{tab:pitfalls}. However, each approach also has its own unique pitfalls, as shown in Table \ref{tab:taxonomy_methods_combined}
. These complexities highlight the need to develop a wide range of guidelines.
Given the broad scale of MI, for explanatory purposes, we explain technical reasons that lead to some pitfalls for one approach: \emph{Activation Patching for Circuit Discovery} \citep{ROME2022}. 

We define model $M$ as a directed graph whose nodes correspond to internal components (e.g., attention heads), and whose edges represent information flow. A \emph{circuit} $C \subset M$ is a subgraph hypothesized to implement a specific behavior, such as induction. We define a metric $F_D(M, C)$ to compare the behaviors of $C$ to $M$ on dataset $D$.

Activation patching replaces a clean prompt into a corrupted one to measure how much performance is restored. Let $a^{\text{clean}}_k$ and $a^{\text{corr}}_k$ be the activations at component $k$, and define a patched model $M^{(k)}$ by replacing only that component:
\[
M^{(k)}_{\text{patched}}(x_{\text{corr}}) = M\bigl(x_{\text{corr}} \ \text{with} \ a^{\text{clean}}_k \rightarrow a^{\text{corr}}_k \bigr).
\]
A typical score then compares a metric $F$ (such as a logit-difference) before and after patching:
\[
\Delta_F(k) = F\bigl(M^{(k)}_{\text{patched}}(x_{\text{corr}})\bigr) - F\bigl(M(x_{\text{corr}})\bigr).
\]
Large values of $\Delta_F(k)$ are usually interpreted as evidence that $k$ contributes to the behavior associated with the clean prompt. This is applied over multiple components to localize a circuit.

Hypothesis-testing formulations note that such scores alone do not ensure that a proposed circuit is well-defined \citep{shi2025hypo}. Even if $\Delta_F(k)$ is large for all $k$ in a candidate subgraph $C$, one must also check that each component is necessary. This is often expressed as a \emph{minimality} condition:
\[
\forall\, k \in C:\quad F_D\!\bigl(M,\,C \setminus \{k\}\bigr) > F_D(M, C),
\]

Because small choices in metrics or circuit reduction can produce wildly different attributions that look equally plausible, practitioners can be mislead into overconfidence in their claims. Corruption schemes can push representations off-distribution, producing spurious $\Delta_F(k)$. Metric selection (e.g., logits vs.\ probabilities) can flip the ranking of important components. 
Therefore, given the sensitivities of experimental conduct to study quality, it is paramount to standardize nuanced guidelines, especially as methods become more complex.

\begin{table*}[t]
\centering
\renewcommand{\arraystretch}{1.15}
\begin{tabular}{|p{4.5cm}|p{5cm}|p{5cm}|}
\hline
\textbf{Pitfall} & \textbf{Description} & \textbf{Auditing Guideline} \\
\hline
Interpretability illusions \citep{Friedman2024} & Plausible explanations that fail when evaluated on broader or counterfactual data. &
Stress-test with counterfactual inputs and hold-out distributions \\
\hline
Cherry-picking \citep{Casper2023EIS} & Demonstrating results only on favorable tasks or hand-picked examples. &
Evaluate on random or full data slices and report negative cases \\
\hline
Missing sanity checks \citep{Adebayo2018} & Failing to test against random models, scrambled labels, or weight shuffling. &
Run method on null baselines and verify degradation \\
\hline
No causal validation \citep{mueller_quest_2025} & Relying solely on correlations or descriptive claims without interventions. &
Use ablations, activation patching, or necessity/sufficiency tests \\
\hline
\end{tabular}
\caption{Examples of potential pitfalls in MI experiments, and guidelines to audit whether experiments avoid them. Auditing checks if these guidelines are followed. Currently, these guidelines are high-level; we expect that as the field progresses and notices patterns in what works, it will develop more nuanced guidelines. More examples are given in Table \ref{tab:more_pitfalls} in Appendix \ref{app:more_guidelines}.}
\label{tab:pitfalls}
\end{table*}

\begin{table*}[h]
\centering
\begin{tabular}{|p{4.5cm}|p{5cm}|p{5cm}|}
\hline
\textbf{Approach} & \textbf{Potential Pitfalls} & \textbf{Auditing Guideline} \\ 
\hline

\textbf{Probing}: Test whether a concept is encoded in activations
& Correlation mistaken for causation \citep{belinkov2021probing}
& Follow with causal interventions \\
\hline

\textbf{Activation Patching}: Test causality via interventions
& Corrupted prompt choice changes outcomes \citep{zhang2024towards}
& Use multiple prompt distributions and track outcome variations \\
\hline

\textbf{Sparse Decomposition}: Decompose polysemantic activations \citep{elhage2022superposition} into interpretable features
& Claim a feature represents a concept, yet may fail to fire on some cases due to absorption \citep{chanin2025absorption}
& Validate coverage by checking recall, and use causal intervention to verify that failures are not due to feature absorption \\
\hline

\textbf{Activation Steering}: Manipulate activations to steer output
& Steering may degrade under OOD distribution shifts \citep{tan2025analyzinggeneralizationreliabilitysteering}
& Evaluate steering on diverse benchmarks \\
\hline
\end{tabular}
\caption{Approach-Specific Auditing Guidelines. These are high-level guidelines that are well-known; however, more complex scenarios will call for more nuanced guidelines that the community should be aware of. These issues may be grouped together under a standardized taxonomy (e.g., some are types of interpretablity illusions).}
\label{tab:taxonomy_methods_combined}
\end{table*}

Additionally, we present an example of using guidelines to audit an MI study in a safety-critical medical scenario in Appendix \ref{app:auditEx}. These examples illustrate that evaluating MI experiments often requires careful inspection of methodological details that are not always visible from headline results alone. While experienced practitioners may recognize such issues informally, there is currently no standardized process for recording, comparing, and auditing these practices across studies, which is essential for AI safety applications described in Appendix \ref{app:impacts}.

\section{Call for Community-Driven Empirical Guidelines}
\label{sec:standards}

To ensure a clear understanding of MI study quality, it is essential to develop effective guidelines. However, given that MI is an emerging field, what constitutes as ``good standards" is still being formulated, and lacks widespread agreement. But if no effort is being made to establish expert-recommended guidelines, the field risks becoming disorganized, confused, and overwhelmed by practices of uncertain quality, such that time and resources would be spent on studies which do not reach their full potential due to a lack of guidance.


Thus, to facilitate the development of effective auditing guidelines, we propose an organized effort to set up a \emph{Collaborative Meta-Analysis Platform} consisting of: (1) Experiment repositories from which the community can organize, review, and cross-compare meta-results, and (2) Forums where users can propose and debate both claims and community-driven ``living document" guidelines akin to Wikipedia\footnote{https://en.wikipedia.org/wiki/Living\_document}, supporting positions with repository-hosted evidence and structured hypotheses chains. 

This platform design assumes that certain guidelines should be backed not just by conceptual justifications, but by real-world empirical evidence. We contend that such an open ecosystem will help ensure that standards are supported by transparent and evidence-based justifications, rather than being established by potentially corruptible authorities based on opaque reasons. By allowing guidelines to be open to revision, we avoid locking in assumptions too early in the field's development.


\subsection{Motivation for Community-Driven Guideline Development}

\noindent\textbf{Current Landscape of MI.}
This paper is not a criticism of how the MI community currently operates; we believe that it is beneficial for a new field to start off as free-form and flexible. At present, it is still unclear what counts as ``good experiment practices", and so innovative exploration should be the norm. However, after years of progress and many successful results, we believe that enough experience has been accumulated for the field to begin moving towards instituting experiment standards. 

Individual MI researchers and organizations have produced a range of valuable educational resources, including well-written explanations of good experiment practices
\citep{nanda2025howtobecome}. Publically-available training curriculum like ARENA have systematized empirical methods.\footnote{https://www.arena.education/team} 
These resources have been very helpful for teaching practioners how to properly conduct MI research. However,
they do not address the systematic auditing of experimental studies. Therefore, we propose that additional resources are needed to more explicitly lay out good experimental practices, both to guide researchers during a project, and to enable systematic post-hoc auditing.

\noindent\textbf{The Open and Online Communities of MI.}
Many advances in MI research have been driven by its open and collaborative online communities \citep{saphra2024mechanistic}, such as forums\footnote{http://lesswrong.com/} and Discord servers\footnote{https://discord.gg/cMr5YqbU4y} where users can engage with MI experts. Additionally, MI research is highly accessible, as it can be done with fewer resources in comparison to other fields. This makes MI share traits with open source development. 
As such, we claim that MI is suited towards utilizing an open experiments platform, as it is not a discipline that relies heavily on closed academic or industry structures. Given that the validities of MI practices are already frequently discussed in these online venues \cite{Casper2023EIS}, organizing these discussions in a user-friendly platform with added functionality is a natural and helpful extension. 

This platform design partly resembles ARBOR, an open collaborations project by Bau Labs (a leading MI lab) in which users can post MI research questions for reasoning models and find collaborators, sharing partial results on experiment pages.\footnote{https://github.com/ARBORproject/arborproject.github.io} Our platform differs in that it is focused on study verification and guideline development. It also shares principles with the MI method leaderboards made by \citet{mueller2025mib}.

\noindent\textbf{Educating a Wider Audience.}
Another benefit from clear guidelines is to make MI more understandable for outsiders and newcomers, allowing academic reviewers who are unfamiliar with the field to use these guidelines as references. Over time, they can potentially evolve and crystallize into canonical textbooks or reference manuals that can be taught in classes at accredited institutes.

\noindent\textbf{Precedents from Established Disciplines.} We base the effectiveness of standardized guidelines on precedents from several fields. In biology, MIAME specifies the minimal reporting conventions required for microarray experiments to be interpretable, reproducible, and comparable \citep{brazma2001miame}. In clinical practices, the GRADE framework, created by a collaborative of methodologists and clinicians, assesses the quality and certainty of medical studies \citep{PRASAD2024101484}. In software engineering, High Integrity C++ is a set of programming guidelines for enforcing reliable, maintainable C++ code in automotive, aerospace, and embedded systems \citep{basalaj2013highIntegrityCpp}.

\subsection{Principles for Guideline Creation}

To ensure guideline quality, the community would benefit from developing meta-guidelines on how to create guidelines. We provide examples as follows:

\noindent\textbf{Guideline Structure. } The structure of a \emph{guideline} is flexible, ranging from a short statement to guides with multiple pages.

\noindent\textbf{Minimal Guidelines.} We propose that these guidelines should not enforce overly rigorous and unjustifiably strict standards. Rather, we advocate that they should only clearly define the minimal requirements that an experiment should adhere to, and allow flexibility in other areas. Personal ``arbitrary” preferences that are not strongly and logically justified should be avoided.

\noindent\textbf{Guides, not Doctrines.} We advocate that these guidelines should not serve as definitive checklists of correctness, but should be used to help practitioners ensure that they are not missing essential elements in their experiments, and to provide more concrete standards for auditors who may not know what to look for.

\noindent\textbf{Encourage Evolution.} We expect that an audit system will take time to mature. It should not be adopted or trusted until it has been tested and refined in practice. Even after refinement, the guidelines will be subject to change.

\section{A Collaborative Platform for Continuous Reviewing}
\label{sec:repository}


\subsection{A Collaborative Meta-Analysis Platform}

We propose establishing a community-driven platform that allows users to post any result, claim, or hypothesis which others can comment on and continue. This approach differs from archives \citep{arxiv, paperswithcode, figshare, f1000research} and journals that collect reproductions \citep{yildiz2020reproducedpapers} or negative results \citep{jinr} as it aims to cultivate a community to continuously interact with and check results. It may be viewed as extending LessWrong with added functionalities, described in this Section and Appendix \ref{app:platform}. As such a platform would not be as useful without active engagement, we call for a community effort to encourage participation. 

\noindent\textbf{Motivation. }
Many \emph{meta-analysis} discussions which take place outside of a paper are highly useful for verifying existing work. 
These discussions and post-hoc results (e.g., replications, reproductions, small counter-results) contain \textbf{meta-knowledge} about papers, including small comments and results which do not fit the paper requirements, but are still highly valuable. We discuss these with more detail in Appendix \ref{app:meta_results} with examples of useful meta-analysis comments.

We suggest that this meta-analysis is a form of continuous reviewing, and already exists in blog posts, Twitter threads, forum posts, private correspondances between researchers, and more. 
However, its meta-knowledge is often buried, scattered, and not dispersed efficiently to others in a community, staying within those who follow or are in correspondence with researchers who can share meta-knowledge. For instance, replies on Discord may be lost within a server, and Twitter threads, forum posts, and servers are at risk of deletion. Meta-knowledge is also often disorganized and hard to parse. For instance, Twitter threads are not optimally designed for scientific discussion, and their content is questionable if those who control them are compromised. These factors make it inefficient to learn good practices for those who are new to the rapidly growing MI community, and LLMs, which may not have been trained on this information, may also be less efficient at finding and using it. 

Therefore, conducting meta-analysis, along with archiving meta-knowledge, in a centralized platform with organizational tools (filtering, recommending, etc.) increases its practical usability. 
We contend that continuously refining meta-knowledge is essential for developing auditing guidelines, as described in Section \ref{sec:guidelines}.
Additionally, we propose that meta-knowledge is a type of \textbf{institutional memory}, which is important community knowledge that is often inadequately documented, and is at-risk of being lost. Previous work has argued that it is essential to record insitutional memory \citep{corbett2018}. 

We posit two essential components to optimally leverage meta-knowledge: \emph{(1) Organization to improve information sharing} and \emph{(2) Live Community-Efforts}.
Our view is that this platform should support two main applications:

\noindent\textbf{Use 1: Continuous, Live Reviewing.}
This is complementary to peer review as this platform is not formal enough to serve as official reviewing; rather, it assists in revision. We propose an approach similar to OpenReview with verified experts and anonymous reviewing, with details in Appendix \ref{app:expert_filtering}.

\noindent\textbf{Use 2: Decentralized Collaboration.}
This platform would allow researchers to not just comment on claims and results, but to share any experiments (partial, post-hoc, etc.) uploaded in \textbf{experiment repositories} with collaborative features similar to GitHub. These repositories host hypotheses, evidence, and claims, along with links to code and papers.\footnote{Similar to https://huggingface.co/papers/trending, but with many more reviewing-centered functionalities.}  
Researchers can comment on repositories, and collaborate to tweak ideas and verify details. 
They can continue partial results that others stopped, were stuck on, or requested help on, due to reasons such as a lack of time. Through iterative contributions, multiple repositories may gradually bring about peer-reviewed papers, and the platform tracks all the (known) contributors to specific results, incentivizing platform engagement.

\subsection{Encouraging Incentives to Engage in Meta-Analysis}
\label{app:1_subsec_reviewer_driven}

While some researchers engage in meta-analysis discussions, reviews, and research, we contend that 
there currently is a lack of strong incentive for others to engage in this practice. 
Given that it can take time, effort, costs, and resources to conduct meta-analysis, and authors are under deadlines to optimally maximize publishing novel papers, engaging in "cleaning work" would comparatively result in little prestige, community engagement, or career-building outcomes, especially on checking "small parts of papers". However, we argue that because such work influences auditing practices, it is important to strengthen incentives to increase engagement, and claim that this platform has the potential to create several strong, new incentives:

\begin{enumerate}
    \item \textbf{Building a Reviewer Portfolio:} This platform may provide user-profile driven incentives for reviewing, such as reviewers who strive to be "expert reviewers" in certain areas, similar to users on Stack Overflow\footnote{https://stackoverflow.com/questions}. Users can review any work, mainly those that are relevant to them, and build a portfolio of reviews. One incentive is that this can act as a resume if it becomes an acceptable standard, similar a GitHub portfolio. The reviews shown publically on a profile are the non-anonymized ones, or those that are later non-anonymized. For non-anonymized posts, a user can click on a profile to see an user's works, which includes experiments they have run, their personal views on claims, papers, and more.
    \item \textbf{Building a Meta-Analysis Portfolio:} Having a platform to share any result may incentivize researchers to undertake useful projects that would not result in a paper, such as negative results that suggest not taking a certain approach \citep{karl2024positionembracingnegativeresults}. As the platform would have features that facilitate partial contibutions, it may encourage more non-paper/partial results to be seen and continued. 
    \item \textbf{Becoming a Partial Contributor:} This platform can utilize a contributor assignment system similar to GitHub, such as showing reproductions and replications a researcher contributed to, negative results they pursued that did not result in a paper, and small corrections to a work they assisted with.
\end{enumerate}

Currently, there are incentives for users on a platform such as LessWrong to engage in online meta-analysis discussion, as these blog posts and comments are shared within the community. However, there is still room for improvement on how this meta-knowledge is utilized. We see this positive engagement on LessWrong as a precedent supporting the view that community-based meta-analysis for MI is practical and valuable, and suggest that organizing these discussions into a platform more optimized for meta-analysis research will increase this engagement even further, especially for newcomers or those who are not part of the community. For instance, there can be pages dedicated to certain claims (e.g., the Linear Representation Hypothesis \citep{park2024linearrepresentationhypothesisgeometry, sutter2025nonlinearrepresentationdilemmacausal}) which collects links to these discussions, and more importantly, encourages continued engagement with a wider audience via features such as recommendation systems on this particular topic, sharing them as "trending topics" that capture more attention. 

\textbf{Incentives Under Anonymity.} This experimental approach can test various anonymity options to assess their effectiveness and flaws. It is possible to still improve meta-analysis engagement for users who are fully anonymous (with no link to a profile) or under a pseudonym (an profile that is not linked to public identity). For instance, this platform can provide a wider and more engaging community that they otherwise would not find on other platforms, and it can provide organizational tools, recommendation systems that suggest work they would be interested in, and more which improve their meta-analysis and reviewing experience, even if the community is fully or partially anonymous. 

\subsection{How a Collaborative Platform Establishes Auditing Guidelines and Protocols}
\label{sec:guidelines}

Both professional auditing and reviewing platforms can falter if its users are not aware of good reviewing practices. Thus, it is crucial to crystallize meta-knowledge about standardized reviewing practices, guidelines, and protocols. We propose a discussion system in which communities work together to generalize common, useful and expert-supported points developed on this platform into guidelines to improve auditing efficiency, both in professional, regulatory practices and on the platform. We suggest for this platform to allow researchers to create \emph{Proposed Guideline} pages where they debate with others over their validity, justifying arguments with empirical sources.

By organizing and connecting experiments together in a shared space, this platform can help the community identify and refine good practices by looking for common patterns across studies, combining their insights into practical auditing guidelines which inform governance and regulation. Additionally, waiting for official ``next editions" of guidelines is a slow process, while MI is a rapidly evolving field; hosting ``living document" guidelines that can be quickly updated can accelerate innovation. 

\begin{figure}[ht]
  \centering
  \includegraphics[width=\columnwidth]{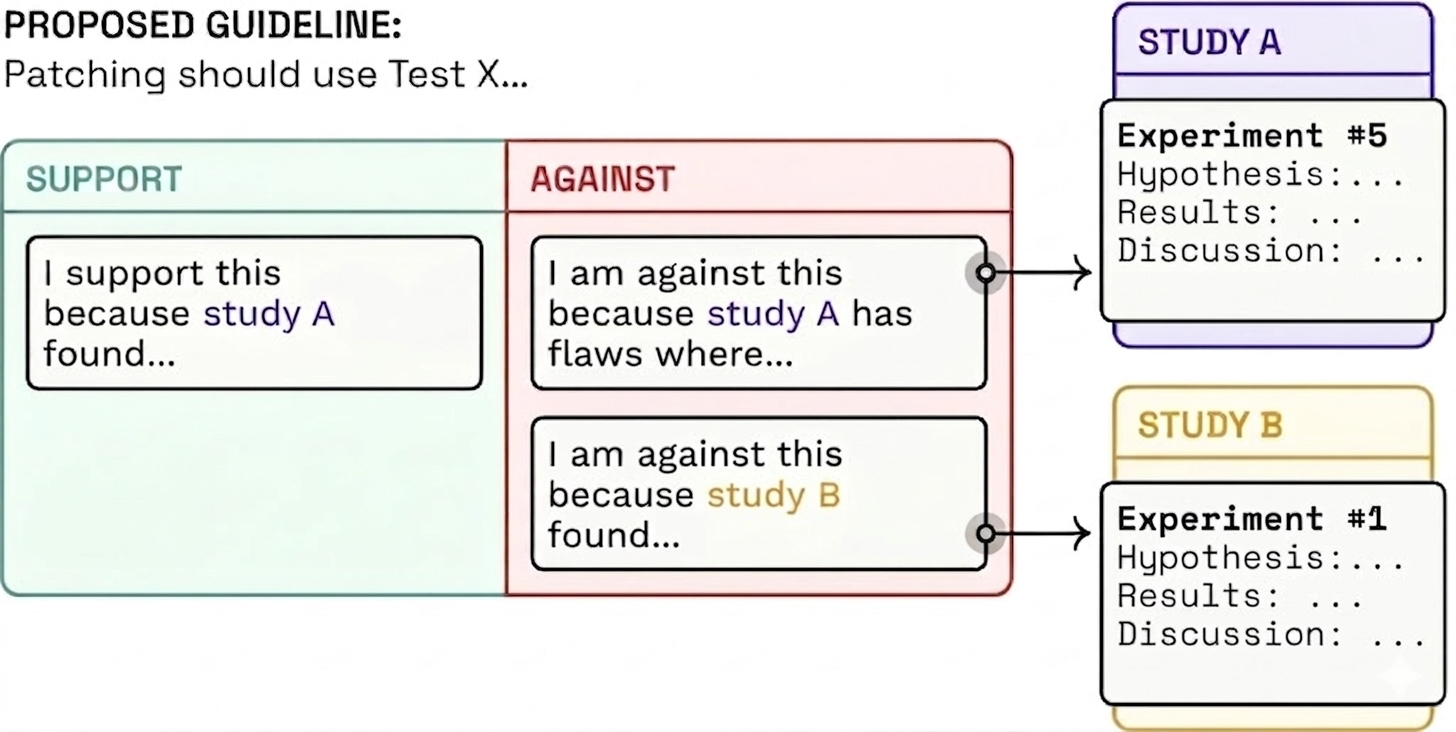}  
  \caption{Example for how proposed guideline discussion pages organize arguments supported by community-reviewed experiment repositories.}
  \label{fig:guideline_page}
\end{figure}

Figure \ref{fig:guideline_page} shows an example of how these guideline discussion pages are formatted, with users posting evidence-based arguments supported by community-reviewed experiment repositories. 
We describe stages on how these freeform discussions lead to standardized, adopted guidelines:
\begin{enumerate}
    \item Users suggest \emph{Proposed Guidelines} pages with arguments supported by empirical evidence, citing papers, repositories, and more. For instance, a Guideline page may propose "Use Test $X$" because previous discussions by researchers found that running Test $X$ from Study $A$ was highly important to assess circuit validity.
    \item On each \emph{Proposed Guideline} page, the community debates about the guideline's validity. For instance, researchers may be against the guideline due to crucial flaws in Study $A$, as discovered by a small meta-result. This may be further supported by Study $B$ findings.
    \item When a professional auditing system seeks to select standardized guidelines, they can consult the arguments on the Proposed Guideline page as evidence-backed references. This ensures that these important meta-analysis discussions are not lost, organizing them in one place. If the auditing system is to be updated, they can track updates via the platform.
\end{enumerate}

We provide high-level examples of guidelines in Appendix \ref{app:more_guidelines}. 
To assist with organzed cross-study comparisons, these experiment formats may benefit from being built on a unified protocol language. We suggest that the community work towards building a user and model friendly protocol which includes formalizing hypothesis testing of causal explanations, and describe this in Appendix \ref{app:protocols}.


\section{Source-Based Automated Auditing}
\label{sec:auto}




\noindent \textbf{Source-Based Reasoning.}
To assist with uncovering claim validity, we suggest developing tools which trace the assumptions, evidence, and other claims which claims depend on. This means that claims discussed on the platform are scrutinized more carefully based on the other statements they depend on, allowing one to trace back statements and evaluate them recursively. This is more specific than citations as this does not just reference papers, but the specific content within them, like plots. We discuss this further in Appendix \ref{app:1_subsec_claims}.

\noindent \textbf{Automated Auditing Assistance via Agentic AI.}
One issue is that there are too many claims for manual source assessment. Thus, we propose the development of automated systems to assess claims at scale that aim to be more objective, possibly using the guidelines curated by the MI community. This approach becomes more important as more automated interpretability methods are developed to accompany human-driven analyses.
This automated system may be built with agentic systems that help trace back long dependency chains of claims, while humans manually verify their outputs to check for and fix issues like hallucinations. 
These agentic systems may also run the code from papers and experiment repositories via evaluation harnesses. 
Additionally, this system may benefit auditors by converting explanations into testable claims with explicit priors that an independent party can recompute. 


\noindent \textbf{Scoring Claims via Probabilistic Logic.}
To assist with how reasons affect claims, researchers and AI agents may find benefits with using probabilistic logic to weigh claims based on claims assumptions (Appendix \ref{app:1_subsec_claims} that discusses this using "Claim Views"). This provides quantifiable estimates of a claim's validity, assisting researchers in their personal understanding of each claim's certainty. However, we note that such a system only suggests values for humans to ultimately verify. 

Overall, this system weighs the strength of a claim by the reasons it depends on to be valid. For instance, if a claim previously depended on assumption $X$, and assumption $X$ was found to be false, then the claim is automatically updated to be very dubious, and a warning alert is sent to those who follow this claim.

We propose an automatic evidence-weighing auditing system be built using rigorous logical probabilistic verification frameworks such as Probabilistic Soft Logic (PSL) \cite{psl_1}, in which hypotheses and observations are linked by logical relations, and weighted rules can be automatically updated such that the likelihood of evidence is maximized, reducing the burden on researchers to hand-tune assumptions.
This system could support handling audit questions such as whether hypotheses were preregistered or post hoc, how sensitive conclusions are to ablations or counterfactuals, exposing when conclusions rest on selective evidence, settling competing explanations, and mitigating the influence of implicit biases.
In Appendix \ref{app:autoaudit}, we discuss the advantages of systems like PSL for auditing, and give an example of a use case for PSL on auditing a study claiming circuit minimality.
We also illustrate a high-level example of automated auditing using a simple LLM based auditing system in Appendix \ref{app:minimal_llm}.



\section{Alternative Views and Considerations}
\label{sec:counters}

We describe two alternate views below and discuss three more in Appendix \ref{app:counters} which address: (i) Guideline definition difficulty, (ii) Being too early to define standards, and (iii) If the field is evolving too fast to have standards.

\noindent \textbf{View 1: Why verify using defined guidelines? Aren’t practical outcomes sufficient?} 
One concern is that practical effectiveness is sufficient evidence of correctness. For instance, if a model steers well, then it ``works”. However, we still have to define what ``steering well” means, as without these more rigorous definitions, it is subject to subjective human interpretation. Essentially, these guidelines aim to ensure that experimental conclusions mirror real-world effectiveness as best as possible. Steering well should be defined based on outcomes of recorded empirical data. It should be noted if past steering experiments conducted on a certain benchmark were representative enough after those approaches were deployed in safety-critical contexts. If it was found that they were not, and that this benchmark rested on brittle assumptions or missed untested edge cases, it is paramount, as a guideline, to re-test the approach on additional benchmarks that better reflect real-world distributions.

\noindent \textbf{View 2: Standardization is restrictive and discourages adoption.} 
Another concern is that formal standards could stifle creativity or impose arbitrary requirements that do not generalize across different research agendas. If researchers do not immediately perceive value in such guidelines, they will be reluctant to adopt such a framework.

However, we do not propose to develop guidelines which are unjustified and strict, but rather, are lightweight, minimal, and necessary enough to just cover what is needed for high quality experiments.
Each guideline should be logically and empirically justified; if a standard is considered arbitrary and not generalizable, the community should encourage its removal.
This leaves substantial room for creative methodologies and novel hypotheses. 

Given that these guidelines aim to be highly useful in strengthening the validity of competing claims, we expect that researchers would see value and demand in this framework.

\section{Conclusion}
\label{sec:conclusion}

Mechanistic interpretability has produced valuable insights, but without standardized auditing procedures the field lacks reliability in safety-critical deployment and governance. Regulatory frameworks increasingly mandate behavioral transparency, and post-hoc explainability methods can greatly improve truth in AI used in financial services, healthcare, and insurance sectors.
We advocate for the reseach community to pursue the development of continuous reviewing, a collaborative approach that gradually refines pieces of research using \emph{meta-analysis} results and discussions that fit outside of a paper. 
We propose that this development will benefit via improved organization of meta-analysis work and discussions using community-driven platforms.
Establishing auditing guidelines arising from community-driven meta-analysis that gradually critiques and refines work outside of papers can greatly assist in refining interpretability for trustworthy AI.







\section*{Limitations}
\label{sec:limitations}
Our position advocates for a goal to establish standardized empirical guidelines for MI, which requires substantial community coordination. We posit an experimental reviewing approach via a platform that depends on active user engagement, which we have not gathered yet, but aim to test in future work, possibly via surveys and workshops. This position paper seeks to obtain this user engagement by encouraging researchers to share their opinions on the necessity and effectiveness of our proposed direction for the MI community, and on how well it can produce standardized guidelines for auditing. Without catalyzing these discussions, there would be fewer users and less attention given to testing this experimental approach in initial trials. Additionally, for real-world guideline adoption by regulatory bodies, key challenges include defining governance structures for guideline development, balancing minimal requirements with methodological flexibility, establishing enforcement mechanisms without stifling innovation, and scaling verification to frontier models where exhaustive testing becomes computationally prohibitive. We do not claim to resolve these tensions fully, as this paper aims to catalyze discussion, rather than to impose solutions.

\bibliography{bibliography}

@misc{eu_ai_act_2024,
  title={Regulation (EU) 2024/1689 of the European Parliament and of the Council of 13 June 2024 laying down harmonised rules on artificial intelligence (Artificial Intelligence Act)},
  author={{European Parliament and Council of the European Union}},
  journal={Official Journal of the European Union},
  year={2024},
  month={July},
  note={Published July 12, 2024}
}

@article{heap2025sparse,
  title={Sparse autoencoders can interpret randomly initialized transformers},
  author={Heap, Thomas and Lawson, Tim and Farnik, Lucy and Aitchison, Laurence},
  journal={arXiv e-prints},
  pages={arXiv--2501},
  year={2025}
}

@article{milliere2026anthropocentric,
    author = {Millière, Raphaël and Rathkopf, Charles},
    title = {Anthropocentric Bias in Language Model Evaluation},
    journal = {Computational Linguistics},
    volume = {52},
    number = {1},
    pages = {379-388},
    year = {2026},
    month = {03},
    issn = {0891-2017},
    doi = {10.1162/COLI.a.582},
    url = {https://doi.org/10.1162/COLI.a.582},
    eprint = {https://direct.mit.edu/coli/article-pdf/52/1/379/2567975/coli.a.582.pdf},
}

@article{mcgrath2023hydra,
  title={The hydra effect: Emergent self-repair in language model computations},
  author={McGrath, Thomas and Rahtz, Matthew and Kramar, Janos and Mikulik, Vladimir and Legg, Shane},
  journal={arXiv preprint arXiv:2307.15771},
  year={2023}
}

@article{nainani2024adaptive,
  title={Adaptive circuit behavior and generalization in mechanistic interpretability},
  author={Nainani, Jatin and Vaidyanathan, Sankaran and Yeung, AJ and Gupta, Kartik and Jensen, David},
  journal={arXiv preprint arXiv:2411.16105},
  year={2024}
}

@article{song2025featureconsistency,
  title={Position: Mechanistic interpretability should prioritize feature consistency in saes},
  author={Song, Xiangchen and Muhamed, Aashiq and Zheng, Yujia and Kong, Lingjing and Tang, Zeyu and Diab, Mona T and Smith, Virginia and Zhang, Kun},
  journal={arXiv preprint arXiv:2505.20254},
  year={2025}
}

@misc{cfpb_circular_2023_03,
  title={Consumer Financial Protection Circular 2023-03: Adverse action notification requirements and the proper use of the CFPB's sample forms provided in Regulation B},
  author={{Consumer Financial Protection Bureau}},
  year={2023},
  month={September},
  day={19},
  url={https://www.consumerfinance.gov/compliance/circulars/circular-2023-03-adverse-action-notification-requirements-and-the-proper-use-of-the-cfpbs-sample-forms-provided-in-regulation-b/}
}

@misc{cfpb_supervisory_highlights_2025,
  title={Supervisory Highlights: Advanced Technologies Special Edition, Issue 38 (Winter 2025)},
  author={{Consumer Financial Protection Bureau}},
  year={2025},
  month={January},
  day={17},
  note={Winter 2025 Edition}
}

@misc{fda_ai_lifecycle_guidance_2025,
  title={Artificial Intelligence-Enabled Device Software Functions: Lifecycle Management and Marketing Submission Recommendations - Draft Guidance for Industry and FDA Staff},
  author={{U.S. Food and Drug Administration}},
  year={2025},
  month={January},
  day={7},
  note={Draft Guidance, Docket No. FDA-2024-D-4488},
  url={https://www.fda.gov/regulatory-information/search-fda-guidance-documents/artificial-intelligence-enabled-device-software-functions-lifecycle-management-and-marketing}
}

@misc{fda_gmlp_principles,
  title={Good Machine Learning Practice for Medical Device Development: Guiding Principles},
  author={{U.S. Food and Drug Administration and Health Canada and Medicines and Healthcare products Regulatory Agency}},
  year={2021},
  month={October},
  url={https://www.fda.gov/medical-devices/software-medical-device-samd/good-machine-learning-practice-medical-device-development-guiding-principles}
}

@inproceedings{Chughtai2023,
    author = {Chughtai, Bilal and Chan, Lawrence and Nanda, Neel},
    title = {A toy model of universality: reverse engineering how networks learn group operations},
    year = {2023},
    publisher = {JMLR.org},
    booktitle = {Proceedings of the 40th International Conference on Machine Learning},
    articleno = {248},
    numpages = {25},
    location = {Honolulu, Hawaii, USA},
    series = {ICML'23}
}

@article{logic_trinh2024solving,
  title={Solving olympiad geometry without human demonstrations},
  author={Trinh, Trieu H and Wu, Yuhuai and Le, Quoc V and He, He and Luong, Thang},
  journal={Nature},
  volume={625},
  number={7995},
  pages={476--482},
  year={2024},
  publisher={Nature Publishing Group UK London}
}

@article{logic_xu2023logiclm,
  title={Logic-lm: Empowering large language models with symbolic solvers for faithful logical reasoning},
  author={Pan, Liangming and Albalak, Alon and Wang, Xinyi and Wang, William Yang},
  journal={arXiv preprint arXiv:2305.12295},
  year={2023}
}

@inproceedings{logic_2023linc,
  title={LINC: A neurosymbolic approach for logical reasoning by combining language models with first-order logic provers},
  author={Olausson, Theo X and Gu, Alex and Lipkin, Benjamin and Zhang, Cedegao E and Solar-Lezama, Armando and Tenenbaum, Joshua B and Levy, Roger},
  journal={arXiv preprint arXiv:2310.15164},
  year={2023}
}

@article{logic_weng2025autoformalization,
  title={Autoformalization in the Era of Large Language Models: A Survey},
  author={Weng, Ke and Du, Lun and Li, Sirui and Lu, Wangyue and Sun, Haozhe and Liu, Hengyu and Zhang, Tiancheng},
  journal={arXiv preprint arXiv:2505.23486},
  year={2025}
}

@inproceedings{psl_1,
  title={A short introduction to probabilistic soft logic},
  author={Kimmig, Angelika and Bach, Stephen and Broecheler, Matthias and Huang, Bert and Getoor, Lise},
  booktitle={Proceedings of the NIPS workshop on probabilistic programming: foundations and applications},
  pages={1--4},
  year={2012}
}

@inproceedings{Stander2024,
    author = {Stander, Dashiell and Yu, Qinan and Fan, Honglu and Biderman, Stella},
    title = {Grokking group multiplication with cosets},
    year = {2024},
    publisher = {JMLR.org},
    booktitle = {Proceedings of the 41st International Conference on Machine Learning},
    articleno = {1889},
    numpages = {27},
    location = {Vienna, Austria},
    series = {ICML'24}
}

@misc{wu2025unifiedverifiedunderstandinggroupoperation,
      title={Towards a unified and verified understanding of group-operation networks}, 
      author={Wilson Wu and Louis Jaburi and Jacob Drori and Jason Gross},
      year={2025},
      eprint={2410.07476},
      archivePrefix={arXiv},
      primaryClass={cs.LG},
      url={https://arxiv.org/abs/2410.07476}, 
}

@misc{anthropic2025claude,
  title        = {Claude Sonnet 4-5 System Card},
  author       = {{Anthropic}},
  year         = {2025},
  howpublished = {\url{https://assets.anthropic.com/m/12f214efcc2f457a/original/Claude-Sonnet-4-5-System-Card.pdf}},
}

@misc{obeso2025realtimedetectionhallucinatedentities,
      title={Real-Time Detection of Hallucinated Entities in Long-Form Generation}, 
      author={Oscar Obeso and Andy Arditi and Javier Ferrando and Joshua Freeman and Cameron Holmes and Neel Nanda},
      year={2025},
      eprint={2509.03531},
      archivePrefix={arXiv},
      primaryClass={cs.CL},
      url={https://arxiv.org/abs/2509.03531}, 
}

@misc{marks2025auditinglanguagemodelshidden,
      title={Auditing language models for hidden objectives}, 
      author={Samuel Marks and Johannes Treutlein and Trenton Bricken and Jack Lindsey and Jonathan Marcus and Siddharth Mishra-Sharma and Daniel Ziegler and Emmanuel Ameisen and Joshua Batson and Tim Belonax and Samuel R. Bowman and Shan Carter and Brian Chen and Hoagy Cunningham and Carson Denison and Florian Dietz and Satvik Golechha and Akbir Khan and Jan Kirchner and Jan Leike and Austin Meek and Kei Nishimura-Gasparian and Euan Ong and Christopher Olah and Adam Pearce and Fabien Roger and Jeanne Salle and Andy Shih and Meg Tong and Drake Thomas and Kelley Rivoire and Adam Jermyn and Monte MacDiarmid and Tom Henighan and Evan Hubinger},
      year={2025},
      eprint={2503.10965},
      archivePrefix={arXiv},
      primaryClass={cs.AI},
      url={https://arxiv.org/abs/2503.10965}, 
}

@misc{amodei2025urgency,
  author       = {Dario Amodei},
  title        = {The Urgency of Interpretability},
  howpublished = {\url{https://www.darioamodei.com/post/the-urgency-of-interpretability}},
  year         = {2025},
  month        = apr
}

@misc{balsam2024goodfireember,
  author       = {Daniel Balsam and Myra Deng and Nam Nguyen and Liv Gorton and Thariq Shihipar and Eric Ho and Thomas McGrath},
  title        = {Goodfire Ember: Scaling Interpretability for Frontier Model Alignment},
  howpublished = {\url{https://www.goodfire.ai/blog/announcing-goodfire-ember}},
  year         = {2024},
  month        = dec
}

@InProceedings{wu2025dila,
  title = 	 {DILA: Dictionary Label Attention for Mechanistic Interpretability in High-dimensional Multi-label Medical Coding Prediction},
  author =       {Wu, John and Wu, David and Sun, Jimeng},
  booktitle = 	 {Proceedings of the 4th Machine Learning for Health Symposium},
  pages = 	 {1014--1038},
  year = 	 {2025},
  editor = 	 {Hegselmann, Stefan and Zhou, Helen and Healey, Elizabeth and Chang, Trenton and Ellington, Caleb and Mhasawade, Vishwali and Tonekaboni, Sana and Argaw, Peniel and Zhang, Haoran},
  volume = 	 {259},
  series = 	 {Proceedings of Machine Learning Research},
  month = 	 {15--16 Dec},
  publisher =    {PMLR},
  url = 	 {https://proceedings.mlr.press/v259/wu25a.html},
}

@inproceedings{Golgoon_2024, series={ICAIF ’24},
   title={Mechanistic interpretability of large language models with applications to the financial services industry},
   url={http://dx.doi.org/10.1145/3677052.3698612},
   DOI={10.1145/3677052.3698612},
   booktitle={Proceedings of the 5th ACM International Conference on AI in Finance},
   publisher={ACM},
   author={Golgoon, Ashkan and Filom, Khashayar and Ravi Kannan, Arjun},
   year={2024},
   month=nov, pages={660–668},
   collection={ICAIF ’24} }

@article{
sharkey2025open,
title={Open Problems in Mechanistic Interpretability},
author={Lee Sharkey and Bilal Chughtai and Joshua Batson and Jack Lindsey and Jeffrey Wu and Lucius Bushnaq and Nicholas Goldowsky-Dill and Stefan Heimersheim and Alejandro Ortega and Joseph Isaac Bloom and Stella Biderman and Adri{\`a} Garriga-Alonso and Arthur Conmy and Neel Nanda and Jessica Mary Rumbelow and Martin Wattenberg and Nandi Schoots and Joseph Miller and William Saunders and Eric J Michaud and Stephen Casper and Max Tegmark and David Bau and Eric Todd and Atticus Geiger and Mor Geva and Jesse Hoogland and Daniel Murfet and Thomas McGrath},
journal={Transactions on Machine Learning Research},
issn={2835-8856},
year={2025},
url={https://openreview.net/forum?id=91H76m9Z94},
note={Survey Certification}
}

@inproceedings{Adebayo2018,
author = {Adebayo, Julius and Gilmer, Justin and Muelly, Michael and Goodfellow, Ian and Hardt, Moritz and Kim, Been},
title = {Sanity checks for saliency maps},
year = {2018},
publisher = {Curran Associates Inc.},
address = {Red Hook, NY, USA},
booktitle = {Proceedings of the 32nd International Conference on Neural Information Processing Systems},
pages = {9525–9536},
numpages = {12},
location = {Montr\'{e}al, Canada},
series = {NIPS'18}
}

@inproceedings{Friedman2024,
author = {Friedman, Dan and Lampinen, Andrew and Dixon, Lucas and Chen, Danqi and Ghandeharioun, Asma},
title = {Interpretability illusions in the generalization of simplified models},
year = {2024},
publisher = {JMLR.org},
booktitle = {Proceedings of the 41st International Conference on Machine Learning},
articleno = {560},
numpages = {25},
location = {Vienna, Austria},
series = {ICML'24}
}

@article{bereska2024mechanistic,
title={Mechanistic Interpretability for {AI} Safety - A Review},
author={Leonard Bereska and Stratis Gavves},
journal={Transactions on Machine Learning Research},
issn={2835-8856},
year={2024},
url={https://openreview.net/forum?id=ePUVetPKu6},
note={Survey Certification, Expert Certification}
}

@misc{Casper2023EIS,
  author       = {Stephen Casper},
  title        = {The Engineer’s Interpretability Sequence},
  howpublished = {Alignment Forum (series of posts)},
  year         = {2023},
  url          = {https://www.alignmentforum.org/s/a6ne2ve5uturEEQK7}
}

@inproceedings{nanda2023progress,
title={Progress measures for grokking via mechanistic interpretability},
author={Neel Nanda and Lawrence Chan and Tom Lieberum and Jess Smith and Jacob Steinhardt},
booktitle={The Eleventh International Conference on Learning Representations },
year={2023},
url={https://openreview.net/forum?id=9XFSbDPmdW}
}

@inproceedings{tigges2024llm,
title={{LLM} Circuit Analyses Are Consistent Across Training and Scale},
author={Curt Tigges and Michael Hanna and Qinan Yu and Stella Biderman},
booktitle={The Thirty-eighth Annual Conference on Neural Information Processing Systems},
year={2024},
url={https://openreview.net/forum?id=3Ds5vNudIE}
}

@misc{belinkov2021probing,
      title={Probing Classifiers: Promises, Shortcomings, and Advances}, 
      author={Yonatan Belinkov},
      year={2021},
      eprint={2102.12452},
      archivePrefix={arXiv},
      primaryClass={cs.CL},
      url={https://arxiv.org/abs/2102.12452}, 
}

@article{mueller_quest_2025,
	title = {The {Quest} for the {Right} {Mediator}: {Surveying} {Mechanistic} {Interpretability} for {NLP} {Through} the {Lens} of {Causal} {Mediation} {Analysis}},
	issn = {0891-2017},
	url = {https://doi.org/10.1162/COLI.a.572},
	doi = {10.1162/COLI.a.572},
	journal = {Computational Linguistics},
	author = {Mueller, Aaron and Brinkmann, Jannik and Li, Millicent and Marks, Samuel and Pal, Koyena and Prakash, Nikhil and Rager, Can and Sankaranarayanan, Aruna and Sharma, Arnab Sen and Sun, Jiuding and Todd, Eric and Bau, David and Belinkov, Yonatan},
	month = sep,
	year = {2025},
	note = {\_eprint: https://direct.mit.edu/coli/article-pdf/doi/10.1162/COLI.a.572/2554934/coli.a.572.pdf},
	pages = {1--48},
}

@misc{chanin2025absorption,
      title={A is for Absorption: Studying Feature Splitting and Absorption in Sparse Autoencoders}, 
      author={David Chanin and James Wilken-Smith and Tomáš Dulka and Hardik Bhatnagar and Satvik Golechha and Joseph Bloom},
      year={2025},
      eprint={2409.14507},
      archivePrefix={arXiv},
      primaryClass={cs.CL},
      url={https://arxiv.org/abs/2409.14507}, 
}

@misc{tan2025analyzinggeneralizationreliabilitysteering,
      title={Analyzing the Generalization and Reliability of Steering Vectors}, 
      author={Daniel Tan and David Chanin and Aengus Lynch and Dimitrios Kanoulas and Brooks Paige and Adria Garriga-Alonso and Robert Kirk},
      year={2025},
      eprint={2407.12404},
      archivePrefix={arXiv},
      primaryClass={cs.LG},
      url={https://arxiv.org/abs/2407.12404}, 
}

@inproceedings{saphra2024mechanistic,
    title = "Mechanistic?",
    author = "Saphra, Naomi  and
      Wiegreffe, Sarah",
    editor = "Belinkov, Yonatan  and
      Kim, Najoung  and
      Jumelet, Jaap  and
      Mohebbi, Hosein  and
      Mueller, Aaron  and
      Chen, Hanjie",
    booktitle = "Proceedings of the 7th BlackboxNLP Workshop: Analyzing and Interpreting Neural Networks for NLP",
    month = nov,
    year = "2024",
    address = "Miami, Florida, US",
    publisher = "Association for Computational Linguistics",
    url = "https://aclanthology.org/2024.blackboxnlp-1.30/",
    doi = "10.18653/v1/2024.blackboxnlp-1.30",
    pages = "480--498",
}

@article{brazma2001miame,
  author       = {Alvis Brazma and Pascal Hingamp and John Quackenbush and Gavin Sherlock and Paul Spellman and Chris Stoeckert and John Aach and Wilhelm Ansorge and Catherine A. Ball and Helen C. Causton and Terry Gaasterland and Patrick Glenisson and Frank C.\,P. Holstege and Irene F. Kim and Victor Markowitz and John C. Matese and Helen Parkinson and Alan Robinson and Ugis Sarkans and Steffen Schulze-Kremer and Jason Stewart and Ronald Taylor and Jaak Vilo and Martin Vingron},
  title        = {Minimum information about a microarray experiment (MIAME) — toward standards for microarray data},
  journal      = {Nature Genetics},
  volume       = {29},
  number       = {4},
  pages        = {365--371},
  year         = {2001},
  doi          = {10.1038/ng1201-365},
  url          = {https://www.nature.com/articles/ng1201-365}
}

@article{PRASAD2024101484,
title = {Introduction to the GRADE tool for rating certainty in evidence and recommendations},
journal = {Clinical Epidemiology and Global Health},
volume = {25},
pages = {101484},
year = {2024},
issn = {2213-3984},
doi = {https://doi.org/10.1016/j.cegh.2023.101484},
url = {https://www.sciencedirect.com/science/article/pii/S2213398423002713},
author = {Manya Prasad},
}

@misc{basalaj2013highIntegrityCpp,
  author       = {Wojciech Basalaj and Richard Corden},
  title        = {High Integrity C++ Coding Standard, Version 4.0},
  howpublished = {Whitepaper, Programming Research Ltd},
  year         = {2013},
  url          = {https://files.iccmedia.com/pdf/prqa150302.pdf}
}

@misc{nanda2025howtobecome,
  author       = {Neel Nanda},
  title        = {How To Become A Mechanistic Interpretability Researcher},
  howpublished = {AI Alignment Forum},
  year         = {2025},
  month        = {sep},
  url          = {https://www.alignmentforum.org/posts/jP9KDyMkchuv6tHwm/how-to-become-a-mechanistic-interpretability-researcher}
}

@inproceedings{shi2025hypo,
author = {Shi, Claudia and Beltran-Velez, Nicolas and Nazaret, Achille and Zheng, Carolina and Garriga-Alonso, Adri\`{a} and Jesson, Andrew and Makar, Maggie and Blei, David M.},
title = {Hypothesis testing the circuit hypothesis in LLMs},
year = {2025},
isbn = {9798331314385},
publisher = {Curran Associates Inc.},
address = {Red Hook, NY, USA},
booktitle = {Proceedings of the 38th International Conference on Neural Information Processing Systems},
articleno = {2997},
numpages = {29},
location = {Vancouver, BC, Canada},
series = {NIPS '24}
}

@inproceedings{zhang2024towards,
title={Towards Best Practices of Activation Patching in Language Models: Metrics and Methods},
author={Fred Zhang and Neel Nanda},
booktitle={The Twelfth International Conference on Learning Representations},
year={2024},
url={https://openreview.net/forum?id=Hf17y6u9BC}
}

@inproceedings{mueller2025mib,
title={{MIB}: A Mechanistic Interpretability Benchmark},
author={Aaron Mueller and Atticus Geiger and Sarah Wiegreffe and Dana Arad and Iv{\'a}n Arcuschin and Adam Belfki and Yik Siu Chan and Jaden Fried Fiotto-Kaufman and Tal Haklay and Michael Hanna and Jing Huang and Rohan Gupta and Yaniv Nikankin and Hadas Orgad and Nikhil Prakash and Anja Reusch and Aruna Sankaranarayanan and Shun Shao and Alessandro Stolfo and Martin Tutek and Amir Zur and David Bau and Yonatan Belinkov},
booktitle={Forty-second International Conference on Machine Learning},
year={2025},
url={https://openreview.net/forum?id=sSrOwve6vb}
}

@misc{geiger2025causalabstractiontheoreticalfoundation,
      title={Causal Abstraction: A Theoretical Foundation for Mechanistic Interpretability}, 
      author={Atticus Geiger and Duligur Ibeling and Amir Zur and Maheep Chaudhary and Sonakshi Chauhan and Jing Huang and Aryaman Arora and Zhengxuan Wu and Noah Goodman and Christopher Potts and Thomas Icard},
      year={2025},
      eprint={2301.04709},
      archivePrefix={arXiv},
      primaryClass={cs.AI},
      url={https://arxiv.org/abs/2301.04709}, 
}

@inproceedings{ROME2022,
author = {Meng, Kevin and Bau, David and Andonian, Alex and Belinkov, Yonatan},
title = {Locating and editing factual associations in GPT},
year = {2022},
isbn = {9781713871088},
publisher = {Curran Associates Inc.},
address = {Red Hook, NY, USA},
booktitle = {Proceedings of the 36th International Conference on Neural Information Processing Systems},
articleno = {1262},
numpages = {14},
location = {New Orleans, LA, USA},
series = {NIPS '22}
}

@misc{elhage2022superposition,
   title={Toy Models of Superposition},
   author={Elhage, Nelson and Hume, Tristan and Olsson, Catherine and Schiefer, Nicholas and Henighan, Tom and Kravec, Shauna and Hatfield-Dodds, Zac and Lasenby, Robert and Drain, Dawn and Chen, Carol and Grosse, Roger and McCandlish, Sam and Kaplan, Jared and Amodei, Dario and Wattenberg, Martin and Olah, Christopher},
   year={2022},
   journal={Transformer Circuits Thread},
}

@misc{gupta2024interpbench,
      title={InterpBench: Semi-Synthetic Transformers for Evaluating Mechanistic Interpretability Techniques}, 
      author={Rohan Gupta and Iván Arcuschin and Thomas Kwa and Adrià Garriga-Alonso},
      year={2024},
      eprint={2407.14494},
      archivePrefix={arXiv},
      primaryClass={cs.LG},
      url={https://arxiv.org/abs/2407.14494}, 
}

@inproceedings{NEURIPS2023_771155ab,
 author = {Lindner, David and Kramar, Janos and Farquhar, Sebastian and Rahtz, Matthew and McGrath, Tom and Mikulik, Vladimir},
 booktitle = {Advances in Neural Information Processing Systems},
 editor = {A. Oh and T. Naumann and A. Globerson and K. Saenko and M. Hardt and S. Levine},
 pages = {37876--37899},
 publisher = {Curran Associates, Inc.},
 title = {Tracr: Compiled Transformers as a Laboratory for Interpretability},
 url = {https://proceedings.neurips.cc/paper_files/paper/2023/file/771155abaae744e08576f1f3b4b7ac0d-Paper-Conference.pdf},
 volume = {36},
 year = {2023}
}

@inproceedings{NEURIPS2018_dc5d637e,
 author = {Manhaeve, Robin and Dumancic, Sebastijan and Kimmig, Angelika and Demeester, Thomas and De Raedt, Luc},
 booktitle = {Advances in Neural Information Processing Systems},
 editor = {S. Bengio and H. Wallach and H. Larochelle and K. Grauman and N. Cesa-Bianchi and R. Garnett},
 pages = {},
 publisher = {Curran Associates, Inc.},
 title = {DeepProbLog:  Neural Probabilistic Logic Programming},
 url = {https://proceedings.neurips.cc/paper_files/paper/2018/file/dc5d637ed5e62c36ecb73b654b05ba2a-Paper.pdf},
 volume = {31},
 year = {2018}
}

@article{MLN,
author = {Richardson, Matthew and Domingos, Pedro},
title = {Markov logic networks},
year = {2006},
issue_date = {February  2006},
publisher = {Kluwer Academic Publishers},
address = {USA},
volume = {62},
number = {1–2},
issn = {0885-6125},
url = {https://doi.org/10.1007/s10994-006-5833-1},
doi = {10.1007/s10994-006-5833-1},
journal = {Mach. Learn.},
month = feb,
pages = {107–136},
numpages = {30}
}

@misc{rocktäschel2017endtoenddifferentiableproving,
      title={End-to-End Differentiable Proving}, 
      author={Tim Rocktäschel and Sebastian Riedel},
      year={2017},
      eprint={1705.11040},
      archivePrefix={arXiv},
      primaryClass={cs.NE},
      url={https://arxiv.org/abs/1705.11040}, 
}

@misc{gansch2025causalbayesiannetworksdatadriven,
      title={Causal Bayesian Networks for Data-driven Safety Analysis of Complex Systems}, 
      author={Roman Gansch and Lina Putze and Tjark Koopmann and Jan Reich and Christian Neurohr},
      year={2025},
      eprint={2505.19860},
      archivePrefix={arXiv},
      primaryClass={cs.RO},
      url={https://arxiv.org/abs/2505.19860}, 
}

@article{soergel2013open,
  title={Open Scholarship and Peer Review: a Time for Experimentation},
  author={Soergel, David and Saunders, Adam and McCallum, Andrew},
  year={2013}
}

@article{corbett2018,
author = {Corbett, Jack and Grube, Dennis C. and Lovell, Heather and Scott, Rodney},
title = {Singular memory or institutional memories? Toward a dynamic approach},
journal = {Governance},
volume = {31},
number = {3},
pages = {555-573},
doi = {https://doi.org/10.1111/gove.12340},
url = {https://onlinelibrary.wiley.com/doi/abs/10.1111/gove.12340},
eprint = {https://onlinelibrary.wiley.com/doi/pdf/10.1111/gove.12340},
year = {2018}
}

@misc{Meloux2025TheDS,
      title={The Dead Salmons of AI Interpretability}, 
      author={Maxime Méloux and Giada Dirupo and François Portet and Maxime Peyrard},
      year={2025},
      eprint={2512.18792},
      archivePrefix={arXiv},
      primaryClass={cs.AI},
      url={https://arxiv.org/abs/2512.18792}, 
}

@misc{shafer1976mathematical,
  title={A Mathematical Theory of Evidence},
  author={Shafer, Glenn},
  year={1976},
  publisher={Princeton university press}
}

@misc{park2024linearrepresentationhypothesisgeometry,
      title={The Linear Representation Hypothesis and the Geometry of Large Language Models}, 
      author={Kiho Park and Yo Joong Choe and Victor Veitch},
      year={2024},
      eprint={2311.03658},
      archivePrefix={arXiv},
      primaryClass={cs.CL},
      url={https://arxiv.org/abs/2311.03658}, 
}

@misc{méloux2025everythingeverywhereoncemechanistic,
      title={Everything, Everywhere, All at Once: Is Mechanistic Interpretability Identifiable?}, 
      author={Maxime Méloux and Silviu Maniu and François Portet and Maxime Peyrard},
      year={2025},
      eprint={2502.20914},
      archivePrefix={arXiv},
      primaryClass={cs.LG},
      url={https://arxiv.org/abs/2502.20914}, 
}

@misc{sutter2025nonlinearrepresentationdilemmacausal,
      title={The Non-Linear Representation Dilemma: Is Causal Abstraction Enough for Mechanistic Interpretability?}, 
      author={Denis Sutter and Julian Minder and Thomas Hofmann and Tiago Pimentel},
      year={2025},
      eprint={2507.08802},
      archivePrefix={arXiv},
      primaryClass={cs.LG},
      url={https://arxiv.org/abs/2507.08802}, 
}

@misc{SecondLookResearch,
  author       = {{Second Look Research}},
  title        = {Second Look Research},
  howpublished = {\url{https://secondlookresearch.com/}},
}

@misc{karl2024positionembracingnegativeresults,
      title={Position: Embracing Negative Results in Machine Learning}, 
      author={Florian Karl and Lukas Malte Kemeter and Gabriel Dax and Paulina Sierak},
      year={2024},
      eprint={2406.03980},
      archivePrefix={arXiv},
      primaryClass={cs.LG},
      url={https://arxiv.org/abs/2406.03980}, 
}

@article{Semmelrock,
author = {Semmelrock, Harald and Ross-Hellauer, Tony and Kopeinik, Simone and Theiler, Dieter and Haberl, Armin and Thalmann, Stefan and Kowald, Dominik},
title = {Reproducibility in machine-learning-based research: Overview, barriers, and drivers},
journal = {AI Magazine},
volume = {46},
number = {2},
pages = {e70002},
doi = {https://doi.org/10.1002/aaai.70002},
url = {https://onlinelibrary.wiley.com/doi/abs/10.1002/aaai.70002},
eprint = {https://onlinelibrary.wiley.com/doi/pdf/10.1002/aaai.70002},
year = {2025}
}

@article{yildiz2020reproducedpapers,
  title={ReproducedPapers.org: Openly teaching and structuring machine learning reproducibility},
  author={Yildiz, Burak and Hung, Hayley and Krijthe, Jesse H. and Liem, Cynthia C. S. and Loog, Marco and Migut, Gosia and Oliehoek, Frans and Panichella, Annibale and Pawelczak, Przemyslaw and Picek, Stjepan and de Weerdt, Mathijs and van Gemert, Jan},
  journal={arXiv preprint arXiv:2012.01172},
  year={2020}
}

@misc{paperswithcode,
  author = {{Papers with Code}},
  title = {Papers with Code},
  url = {https://paperswithcode.com/},
}

@misc{f1000research,
  author = {{F1000Research}},
  title = {F1000Research: An open research publishing platform},
  url = {https://f1000research.com/},
}

@misc{jinr,
  author = {{Journal of Interesting Negative Results}},
  title = {Journal of Interesting Negative Results},
  url = {https://www.journals.elsevier.com/journal-of-interesting-negative-results},
}

@misc{figshare,
  author = {{Figshare}},
  title = {Figshare: Research data repository},
  url = {https://figshare.com/},
}

@misc{arxiv,
  author = {{arXiv}},
  title = {arXiv e-Print archive},
  url = {https://arxiv.org/},
}

@inproceedings{
wang2023interpretability,
title={Interpretability in the Wild: a Circuit for Indirect Object Identification in {GPT}-2 Small},
author={Kevin Ro Wang and Alexandre Variengien and Arthur Conmy and Buck Shlegeris and Jacob Steinhardt},
booktitle={The Eleventh International Conference on Learning Representations },
year={2023},
url={https://openreview.net/forum?id=NpsVSN6o4ul}
}

@misc{wang2022ioi_lessons,
  title        = {Some Lessons Learned from Studying Indirect Object Identification in GPT-2 Small},
  author       = {Wang, Rowan and Variengien, Alexandre and Conmy, Arthur and Shlegeris, Buck and Steinhardt, Jacob},
  journal      = {AI Alignment Forum},
  year         = {2022},
  note         = {\url{https://www.lesswrong.com/posts/3ecs6duLmTfyra3Gp/some-lessons-learned-from-studying-indirect-object}}
}

@article{lange2023patching_subspaces_illusion,
  title        = {An Interpretability Illusion for Activation Patching of Arbitrary Subspaces},
  author       = {Lange, Georg and Makelov, Alex and Nanda, Neel},
  journal      = {AI Alignment Forum},
  year         = {2023},
  note         = {\url{https://www.lesswrong.com/posts/RFtkRXHebkwxygDe2/an-interpretability-illusion-for-activation-patching-of}}
}

@misc{chan2022causal_scrubbing,
  title        = {Causal Scrubbing: a Method for Rigorously Testing Interpretability Hypotheses},
  author       = {Chan, Lawrence and Garriga-Alonso, Adri{\`a} and Goldowsky-Dill, Nicholas and Greenblatt, Ryan and Nitishinskaya, Jenny and Radhakrishnan, Ansh and Shlegeris, Buck and Thomas, Nate},
  journal      = {AI Alignment Forum},
  year         = {2022},
  note         = {\url{https://www.lesswrong.com/posts/JvZhhzycHu2Yd57RN/causal-scrubbing-a-method-for-rigorously-testing}}
}

@misc{greenblatt2023useful_mech_interp,
  title        = {How Useful Is Mechanistic Interpretability?},
  author       = {Greenblatt, Ryan},
  journal      = {LessWrong},
  year         = {2023},
  note         = {\url{https://www.lesswrong.com/posts/tEPHGZAb63dfq2v8n/how-useful-is-mechanistic-interpretability}}
}

@misc{abraham2025deep_saes_too,
  title        = {Deep Sparse Autoencoders Yield Interpretable Features Too},
  author       = {Abraham, Armaan A.},
  journal      = {AI Alignment Forum},
  year         = {2025},
  note         = {\url{https://www.lesswrong.com/posts/tLCBJn3NcSNzi5xng/deep-sparse-autoencoders-yield-interpretable-features-too}}
}

@misc{farnik2025x_random_saes_discussion,
  author       = {Farnik, Lucy},
  title        = {Great to see so much discussion of our recent paper! Leo's take here is likely true},
  year         = {2025},
  howpublished = {\url{https://x.com/lucyfarnik/status/1885967412375290097}},
}

@misc{marks2025x_randomized_sae_thread,
  author       = {Marks, Samuel},
  title        = {In fact, this experiment has been done at least 3 times!},
  year         = {2025},
  howpublished = {\url{https://x.com/saprmarks/status/1885869042629857590}},
}

\appendix

\newpage
\appendix
\onecolumn

\section*{\LARGE Appendix}
\addcontentsline{toc}{section}{Appendix}

\vspace{1em}

\makeatletter
\newcommand{\appsectionline}[3]{%
  \vskip 0.85em
  \@dottedtocline{1}{0em}{2.6em}{\large\numberline{#1}#2}{\large\pageref{#3}}%
}
\newcommand{\appsubsectionline}[3]{%
  \vskip 0.55em
  \@dottedtocline{2}{2.6em}{3.4em}{\large\numberline{#1}#2}{\large\pageref{#3}}%
}
\newcommand{\appsubsubsectionline}[3]{%
  \vskip 0.25em
  \@dottedtocline{3}{6.0em}{4.3em}{\large\numberline{#1}#2}{\large\pageref{#3}}%
}
\makeatother

\appsectionline{A}{A Continuous Reviewing Platform for Meta-Results: Extended Description}{app:platform}
\appsubsectionline{A.1}{How do we Define Meta-Results?}{app:meta_results}
\appsubsectionline{A.2}{Building a Collaborative Reviewing Culture}{app:1_subsec_collaborative}
\appsubsectionline{A.3}{Filtering based on Expert Opinions}{app:expert_filtering}
\appsubsectionline{A.4}{Tracking Assumptions via Claim Dependency Graph Tools}{app:1_subsec_claims}
\appsubsectionline{A.5}{Benefits of Decentralized Collaboration}{app:1_subsec_benefits_decentralized}
\appsubsectionline{A.6}{Recommendation and Search Systems}{app:1_subsec_recommendation}
\appsubsectionline{A.7}{Organization Tools for Cross-Experiment Analysis}{app:1_subsec_organization_tools}

\appsectionline{B}{A Step-by-Step Example of Practical MI Auditing}{app:auditEx}
\appsubsectionline{B.1}{Example of Auditing a Study for Safety-Critical Medical Scenarios}{app:2_subsec_example_audit}
\appsubsectionline{B.2}{Example of Generalized Auditing Steps for Activation Patching}{app:2_subsec_example_activation_patching}

\appsectionline{C}{More Guideline Examples}{app:more_guidelines}
\appsectionline{D}{More Alternative Views}{app:counters}
\appsectionline{E}{Importance of MI Auditing for Regulatory Policies}{app:impacts}
\appsectionline{F}{MI Protocols}{app:protocols}
\appsectionline{G}{Automated Auditing Assistance using Probabilistic Logic}{app:autoaudit}
\appsectionline{H}{Auditing Minimal-Circuit Claims}{app:minimal_llm}


\twocolumn

\section{A Continuous Reviewing Platform for Meta-Results: Extended Description}
\label{app:platform}







\subsection{How do we Define Meta-Results?}
\label{app:meta_results}

Meta-results contain information that is useful, but often is not found within a paper. These include critiques, comments (e.g., small comments on one part of a paper, even on an appendix study), seemingly minor negative results, replications (conducting the exact same study), reproductions (conducting the study under varied conditions), non-novel extensions (e.g. just add an ablation study, try it in a new framework, etc.), minor results (which, by themselves, are not enough for a paper, but can be useful in future if built on by others), and partial results (e.g., still requires an ablation study). 

Experiment repositories allow hosting findings that are fall outside papers, such as a specific feature/circuit found using existing methods which are not novel enough for a paper, but can useful enough to share, potentially contributing to auditing by finding counter use-cases, or compiled into a database with other minor findings. 

Examples of useful meta-results can be found in comments on platforms such as LessWrong \citep{wang2022ioi_lessons, chan2022causal_scrubbing, lange2023patching_subspaces_illusion, greenblatt2023useful_mech_interp, abraham2025deep_saes_too} and Twitter/X \citep{farnik2025x_random_saes_discussion, marks2025x_randomized_sae_thread}.

For instance, users have posted meta-results in the form of post-hoc re-evaluations or reproductions, such as a comment in a LessWrong post where a researcher revisited IOI \citep{wang2023interpretability} using causal scrubbing \citep{chan2022causal_scrubbing} and found that it held up differently than the original paper’s faithfulness-style metric suggested, with recovered effects that was only fractions of the original logit differences \citep{wang2022ioi_lessons}. They state that there is "uncertainty in interpreting these numbers"; we posit that small, specific open questions like these are useful for passing along to others who can help contribute to them, and propose tools that give them greater visibility. Additionally, these questions may have likely been answered later (e.g., this is an older topic that already had a good amount of discussion), and so archiving how they were answered later is also valuable.

Another example of valuable meta-analysis is a discussion over claims from a paper on randomized SAEs, in which researchers defended their claims, clarified misunderstandings, and shared meta-results \citep{farnik2025x_random_saes_discussion, marks2025x_randomized_sae_thread}. 

However, we note that not all meta-analysis discussions and results are as neatly archived as these LessWrong discussions. In addition, it would be helpful to have a place to showcase and find these results better, such as using organized listings and networks of experiment repositories related to certain claims (e.g., the Linear Representation Hypothesis \citep{park2024linearrepresentationhypothesisgeometry, sutter2025nonlinearrepresentationdilemmacausal}), as this can improve collaborative meta-analysis. Storing meta-results in experiment repositories, instead of just reporting them in comments that are response to blogs, can enhance reproducibility, their use in future work, and their use in informing guidelines.

\subsection{Building a Collaborative Reviewing Culture}
\label{app:1_subsec_collaborative}
To pass down meta-knowledge efficiently, it would be advantageous to cultivate a reviewing culture that encourages sharing information and good practices.
A "Review Hub" platform can allow users to highlight to make comments, like Overleaf or Google Docs, and ask questions if something is unclear. This platform also helps each user personally understand claims better and share their understanding with others. 
Continuous reviewing and commenting would also allow targetting very specific parts of a paper or experiment (e.g., Why is this plot justified? Why use this metric?), rather than requiring users to review entire papers.

Such a platform would not serve as official peer review, given that it is less robust to issues such as bias and quality, but would serve as a drafting and brainstorming tool that assists with collaboratively verifying and improving upon claims. It can also help assess a researcher's own personal beliefs. Thus, the platform does not aim to assign scores to projects/papers. 
Optionally, the platform may have systems which give scores to claims, but these are merely suggestions used in one factor of validation. This helps drafts before they are put in peer review.

\subsection{Filtering based on Expert Opinions}
\label{app:expert_filtering}

One issue with a public system that any user can comment on is that there can be too much information, including spam. Thus, the platform can benefit users by providing them with multiple, customizable options for what information to filter (e.g., hide content with many down votes, show all content without voting influence, etc.) Having multiple options can allow users to assess multiple viewpoints, rather than just locking into potentially biased communities or authoritative figures, while also assessing selective expert opinions to find content that is assumed to be more likely of higher quality.
On project pages, users and authors may filter to only show content from 'expert-verified' profiles. Verification can be done similar to how it is vetted on OpenReview, or may directly import an OpenReview profile.
Though the profile is verified, the identity can be anonymized to the public. Double blind anonymity or non-anonymity options may be allowed.


\textbf{Voting and Automated Filtering via Argument Rigor. }
Another possible way to weigh user consensus for use in voting and filtering is by assessing the strength of their reasoning. For instance, an agentic reasoning LLM can determine if an argument was logically justified, and check if user is accurately representing evidence interpretation in their writing.
Voting on content (to hide, show on top, claim validity, etc.) may also require logical justification, assessed by an LLM-based system, for it to be counted or heavily weighed. This does not have to be the default view, but can be a user-selected option on how to display/rank content.


\subsection{Tracking Assumptions via Claim Dependency Graph Tools}
\label{app:1_subsec_claims}
We propose a system to improve how researchers assess argument validity via "claim dependency lists/graphs". These tools may help with collaborative debates by formatting, organizing, and visualizing large information traces better.
We define \textbf{statements} as pieces of content: phrases, entire papers, a plot, etc. We define \textbf{claims} as statements asserted to be true (to some degree). We propose a tool in which, given a claim, users construct arguments by linking to pieces of other claims, ending at root statements (assumptions) that are subjectively satisfactory for the current debate. 

During debates, a researcher can ask to further unfurl a claim (e.g., ask "what does it depend on?"), and also debate what reasons each statement should actually depend on. One user may propose one reason to support a claim, and another may accept it, or question that reason even further, asking for more scrunity on a certain benchmark it claimed to have passed. 

Essentially, creating a dependency tracking tool allows researchers to keep asking for sources of sources while keeping track of potentially long dependency chains, with each claim node having with multiple proposed reasons supporting them. This may help with recursively questioning assumptions that claims are dependent on, which can become unorganized as one keeps on tracing reasons back.
For instance, a researcher may assume result X from paper P is true. When constructing an argument, researchers may have unquestionably assumed this was true as they did not explicitly unfurl this reason before. However, with this tool, they have be more prone to questioning it, and finding "bugs" they missed before.

Due to the many reasons that claims depend on, and dependency chains which may go on infinitely, no dependency graph is assumed to be exhaustive. During arguments, a researcher's claim depends on certain key reasons; the dependency graph only needs to be made to contextually be useful in that specific debate. They can refer to specific areas of some piece (e.g., a specific plot). Then, reviewers hone in on that plot and debate the veracity of that plot, since its validity contributes to the claim.

Thus, a claim graph is not exhaustive; it is an organizational scratchpad tool used to aid in discussions over claim validity. As such, it only displays a select number of claims which a claim depends on in the given context. 
Given one important claim which the paragraph is centered on defending, we create a graph of claims which the claim depends on. From there, we can further unfurl each claim node into possible reasons to support them. During discussions, a researcher can further add reasons as to why a reason supports a claim, and mark the edge as "weak" or "strong".

We note that claim graphs cannot replace written arguments, as the reasons which support a claim can often be too complex to fully capture in a claim graph. However, they are only meant as an organizational tool to assist in thinking and discussion.

\noindent \textbf{Implementation Details. }
Claims are created by users. Researchers can state "I create this claim and show what it depends on". 
Users can click on a claim to open a UI element which shows what that claim depends on. They can keep on clicking on claims to open up sources of sources. This can be shown as a list or graph visualizer, with the former being easier to use for claims with many dependencies. 

Users can propose a claim in a paper/experiment/claim/guideline page and give what reasons they believe it depends on. In public pages, users vote to keep what proposed claims should be shown. 
Others can subjectively state that a claim still needs more reasons. In the public views, a researcher can see the votes agreeing with this or not (with the option of not having votes in personal views unless turned on by the host). Claims or dependencies that are voted down by a well-calculated metric are hidden or removed to prevent clutter. 

The platform would aim to allow an easy-to-use construction of claim graphs. Claim graphs can be saved onto the platform to assist others in understanding the reasons supporting a claim, and allowing them to suggest improvements, such as by questioning assumptions (e.g., this metric was found to have a flaw).
A researcher can trace back claims of claims to find more relevant pieces to review.

\noindent \textbf{Claim Pages. }
Optionally, some claims can become pages. 
Researchers can create pages that focus on claims themselves, rather than on specific papers. This means researchers focus on the results, and less on authors of specific papers. This allows results from multiple papers to be collected into a more organized and archived discussion, allowing arguments that support claims to be traced more specifically to parts of papers, not just citing entire papers.


\noindent \textbf{Automated Claim Weighing. }
Claims can be assigned validity scores based on other claims they depend on, assigned weight by researchers or automated LLM agents. When an ancestor node's value is updated, these can propagate to update other claim nodes based on the conditional dependence structure.
Note that this is often not objective enough, but just "convincing enough". Hence, we do not have strong conclusive "weighing systems"; the weighing systems are only meant to give suggested, estimated, calculated snapshot scores, similar to statements made by LLMs. It is up to humans to use those scores or not. 

\subsubsection{Claim Dependency Views}
\label{app:1_subsubsec_claim_dependency}
Given that the reasons which claims are dependent upon can be subjective, we suggest the use of Claim Dependency Views (i.e., \textbf{Views}), which are subjective dependency graphs of what claims depend on. Different researchers may disagree on what reasons a claim depends on, and thus can build their own views to share with other researchers, which may improve information organization and ease of collaborative debates.
Users can fork views to continue in their own way, and historically trace where they were built off from. Others can propose changes to a view, which the author can accept.

To prevent spam by bots and trolls, there can be are "verified" views which anyone who is verified can comment on. This helps prevent gatekeeping from experts, but has the downside that of noisier arguments from non-experts. Thus, to filter out the opinions of non-experts, there is can be a "verified expert" View. Verification is done by ensuring a user is a human using identification systems. Experts are verified similar to how conferences select for reviewers, based on reputation and references. In Verified Expert views, experts can only interact in the field or sub-field they are verified in.

Overall, we propose four main types of public Views: General, Verified, and Verified Expert, along with Personal Views. Within each view, one can show them in three different ways: All displays every post without votes, Vote-Filtered displays with Votes, hiding those with very low votes to prevent clutter, and Custom is user-customized (e.g., filter out certain users). 
To prioritize community-weighed quality information, public views can have reasons removed by a sophisticated voting consensus (to be determined) which is designed to be robust against manipulation. For personal views, a user can create as many copies as they want to help with validity reasoning, forming different conclusions under different assumptions.

Each argument can be measured by several metrics, including community votes (both unweighted and reputation-weighted metrics). Users can rank arguments in various ways by filtering for certain metrics. Guidelines may be scored across multiple metrics, including argument certainty (based on validity of prior assumptions, etc.) 

\subsection{Benefits of Decentralized Collaboration}
\label{app:1_subsec_benefits_decentralized}


Researchers can post work that needs to be done or checked, laying out proposed roadmap for others to take up.
Other users can fork and continue repos in their own way.
This helps prevents partial progress or questions from being forgotten, and helps maintain existing work with minor extensions.

\subsection{Recommendation and Search Systems}
\label{app:1_subsec_recommendation}
To fuel community engagement and direct researchers to interact with work they are familiar with, this platform can benefit by using a recommendation system to \textcolor{blue}{relay} and rela\textcolor{blue}{t}e results.
Users can customize what is recommended, and choose claims, experiments, and drafts to review. They can search for and filter by expertize, sub-field, and specific claims they tackle.

\subsection{Organizational Tools for Cross-Experiment Analysis}
\label{app:1_subsec_organization_tools}
Currently, experiment results are recorded in papers or blog posts. These results are enshrouded in dense, unstructured text; it would be beneficial to distill them into more structured formats for easier comparison. 
Organized repositories allow users to more easily search, filter, and cross-compare fine-grained results across studies. Users can search for specific experiments and their variations (e.g., hyperparameter sweeps) within a study, checking their thoroughness. Users can create custom dashboards from which they can cross-compare results, and notice patterns that they can abstract into general guidelines.

\section{A Step-by-Step Example of Practical MI Auditing}
\label{app:auditEx}




\subsection{Example of Auditing a Study for Safety-Critical Medical Scenarios}
\label{app:2_subsec_example_audit}

\noindent\textbf{Setting.} Clinicians want to detect signs of a heart attack from short patient descriptions. They use a model to determine whether patients have signs of a heart attack or not.

\noindent\textbf{Use of MI for fine-tuning / pruning. } When clinical guidelines change (e.g., atypical presentations in women, diabetics, or elderly patients), knowing the internal circuit lets developers fine tune or prune only the relevant components instead of retraining the entire model. Hospitals can also re-audit over time or across subpopulations.

\noindent\textbf{Experiment Claims.} After activation patching, the researchers conclude that specific model components perform the reasoning needed to flag high-risk cases. Their paper presents highly impressive results.

\noindent\textbf{Step by Step Audit.} We describe an audit using activation patching guidelines.
\begin{enumerate}
  \item \textit{Frame the objective.} Auditors define the target behavior precisely. The model must label a short description as urgent when symptoms like chest pressure and radiating arm pain appear.
  \item \textit{Collect Reproducibility Artifacts.} They pin the checkpoint, tokenizer, seeds, hardware, code commit, and the exact evaluation set. This ensures that any divergence is meaningful rather than accidental.
  \item \textit{Read the data.} They inspect a sample of clean prompts and the study’s corrupted counterparts. The corrupted texts replace key symptom phrases with realistic alternatives that alter clinical meaning while preserving grammar.
  \item \textit{Test the corruption.} They verify that corruption produces a measurable drop in correct urgent labeling, yet keeps text fluent and in distribution. This confirms that a restoration effect would matter.
  \item \textit{Probe granularity.} They retrace the authors’ path from residual stream to layer to head to token position. Patch effects appear where the paper reports them, which suggests the localization procedure was applied carefully.
  \item \textit{Establish baselines.} They run and archive three states. Clean, corrupted, and patched corrupted. The corrupted state shows degraded performance and the patched state seems to restore it.
  \item \textit{Small scale reproduction.} Using fresh random seeds and equivalent prompts, they reproduce headline plots on a held out slice. Effect sizes are similar, which increases confidence in implementation fidelity.
  \item \textit{Critical metric audit.} The auditors notice a dire issue: \textbf{\textcolor{red}{the paper scores success only by the probability of the urgent label}}. Auditors recompute with \textit{logit difference}, comparing urgent against a non urgent alternative,
    \[
      \Delta \ell = z(\text{urgent}) - z(\text{non urgent}) .
    \]
    Several components that looked important under probability show negligible or even negative \(\Delta \ell\). The probability gains came from incidental shifts in the output distribution, not from genuine causal influence on the decision.
  \item \textit{Decision and remediation.} Since the metric masks the true effect, \textbf{the central claim does not hold}, and the study cannot be used yet for clinical decision-making. The study must be repeated with logit based or KL based evaluations, pre registered metrics, and the same otherwise strong design. Corruption, granularity, and baselines were sound, but the metric choice invalidates the causal conclusion.
\end{enumerate}

Note that just because a study passes one auditing test, does not mean it is automatically robust. For one, there may be aspects to test which that auditing test is missing. However, it does mean that the study is seen as more robust than before the auditing was conducted. Therefore, auditing ascribes more certainty to the study's validity, increasing confidence in its use in safety-critical applications.

\subsection{Example of Generalized Auditing Steps for Activation Patching}
\label{app:2_subsec_example_activation_patching}
\begin{enumerate}
    \item Define the target behavior and hypotheses.
    \item Collect all artifacts: model version, code, prompts, random seeds, and metrics.
    \item Validate prompt construction and the evaluation set for relevance and leakage.
    \item Examine the corruption method; confirm it is in-distribution and causes a measurable behavioral shift.
    \item Verify that the evaluation metric aligns with the hypothesis (e.g., logit difference rather than raw probability).
    \item Confirm that the patch targets and granularity match the causal claim.
    \item Establish clean and corrupted baselines prior to patching.
    \item Reproduce a subset of patch results to confirm the reported effect.
    \item Perform sensitivity analyses with alternative metrics, corruptions, seeds, and prompt distributions.
    \item Review causal interpretation, check for negative or redundant components, and document limitations and recommendations.
\end{enumerate}

On the online platform, users will be able to propose freeform guides such as this example. The community can then discuss, agree with, and critique this item, refining it over time.

\section{More Guideline Examples}
\label{app:more_guidelines}






\begin{table*}[t]
\centering
\renewcommand{\arraystretch}{1.15}
\begin{tabular}{|p{4.5cm}|p{5cm}|p{5cm}|}
\hline
\textbf{Pitfall} & \textbf{Description} & \textbf{Auditing Guideline} \\
\hline
Statistical fragility \citep{Meloux2025TheDS}
& Plausible-looking explanations may not be uniquely identified by the evidence. 
& Test alternative hypotheses and quantify uncertainty across null baselines and perturbed settings \\
\hline
Over-reliance on automated interpretability \citep{heap2025sparse} & Automated labeling methods may produce seemingly interpretable features even in randomly initialized models, suggesting that labels may not correspond to genuine concepts. &
Validate features through causal tests and comparison with random or null models \\
\hline
Feature instability across SAE runs \citep{song2025featureconsistency} & Sparse autoencoders trained with different random seeds can produce different features, making single-run interpretations unreliable. &
Check whether features are consistently recovered across multiple SAE training runs \\
\hline
Ignoring redundant or adaptive circuits \citep{mcgrath2023hydra} & Networks may contain redundant pathways that compensate for ablations, as shown by the Hydra effect, making single-pathway explanations incomplete. &
Test robustness of explanations under multiple ablations and intervention patterns \\
\hline
Anthropomorphic projection \citep{milliere2026anthropocentric} & Human-intuitive narratives may be projected onto model internals without reflecting the model’s actual computation. &
Prefer operational or causal descriptions over anthropomorphic narratives \\
\hline
Sensitivity to prompt format or dataset curation. \citep{nainani2024adaptive} & Circuit behavior can change substantially with prompt format, data distribution, or random seed choices. &
Evaluate findings across multiple prompt templates, datasets, and random seeds \\
\hline
\end{tabular}
\caption{More examples of potential pitfalls in MI experiments, and guidelines to audit whether experiments avoid them. Auditing checks if these guidelines are followed.}
\label{tab:more_pitfalls}
\end{table*}

\textbf{Examples of Guidelines Types. } 
\begin{enumerate}
    \item Guidelines for hypothesis testing
    \item Guidelines for evaluating observations (i.e., do the study's claims match what the methods actually find?)
    \item Guidelines for comparing methodologies
    \item Guidelines for designing benchmarks
\end{enumerate}

\textbf{Examples of Standardized Definitions (for Circuit Discovery):} We base these defintions on the theoretical framework of causal abstraction \citep{geiger2025causalabstractiontheoreticalfoundation}.
\begin{itemize}
  \item \textbf{Hypothesis}: a high-level causal model \(H\) over low-level model internals \(N\)  that posits how inputs, latents, and outputs relate.
  \item \textbf{Feature}: a component in the high-level model \(H\) that carries a specific causal role. 
  \item \textbf{Causal Abstraction}: a mapping \(\tau\) that relates low-level variables in \(N\) to high-level variables in \(H\), preserving (or approximately preserving) causal relationships under permissible interventions. 
  \item \textbf{Faithfulness}: a quantitative measure of how closely the high-level model \(H\) approximates the interventional behavior of \(N\).  
\end{itemize}


\textbf{Examples of Criteria to Test (for Circuit Discovery):}
\begin{enumerate}
    \item \textbf{Behavior Preservation}: Intervening to route model computation through the proposed circuit must preserve the model’s task behavior relative to the unmodified model within a small, predefined tolerance. 
    \item \textbf{Localization}: Perturbations restricted to the hypothesized circuit should reproduce the effect, and perturbations outside the circuit should not. 
    \item \textbf{Minimality}: Remove components that are not necessary without reducing performance on the target behavior beyond a preset tolerance. 
    \item \textbf{Distribution Shift Robustness}: Guarantee that their interpretations can be maintained across distribution shifts in the data.
\end{enumerate}

\textbf{Examples of Benchmark Standards:}
\begin{itemize}
  \item Test on intervention benchmarks like MIB \citep{mueller2025mib} and InterpBench. \citep{gupta2024interpbench} to compare circuit localization and causal variable localization across methods. For instance, MIB provides a benchmark with standardized tasks, counterfactual datasets, principled metrics, and private-test-set leaderboards for both localization and featurization, encouraging more stable cross-method comparison. InterpBench complements this by providing 86 semi-synthetic transformers with known circuits, making the ground-truth causal structure more trustworthy, so researchers can test whether a method actually recovers the right mechanism rather than only producing plausible explanations.
  \item Include stress tests across i.i.d. and out-of-distribution shifts to assess robustness of interpretability claims.  
  \item Enforce replicability checks by requiring that published experiments be rerun with small perturbations (e.g., input noise or weight initialization variation).  
\end{itemize}


\textbf{Examples of Model Organism Standards:}
\begin{itemize}
  \item Declare reference models that are tractable to inspect. 
  \item Require version control of published checkpoints, hyperparameters, and training logs to ensure reproducibility.  
  \item Standardize documentation of dataset provenance, training protocol, random seeds, architecture details.
\end{itemize}

\textbf{Examples of General Practices:}

\begin{itemize}

  \item \textbf{Reporting standard:} Report intervention operators, ablation knobs, thresholds, datasets, and code used to run the tests. Use a structured schema so an auditor can recompute the same statistics.

  \item \textbf{Ground-truth validation where possible:} When a ground-truth mechanism exists, require that methods recover it before application to opaque models. Use Tracr compiled transformers as ``known-mechanism'' testbeds to validate tooling and to calibrate localization and minimality metrics \citep{NEURIPS2023_771155ab}. 

  \item \textbf{Benchmark against community tasks:} Evaluate methods on public, multi-task benchmarks that score both circuit localization and causal variable localization. 

  \item \textbf{Sanity checks and falsification attempts:} Run randomization and model-parameter permutation test to ensure results are model and data-dependent. If a purported circuit explanation survives when labels or weights are randomized, the explanation fails a minimum bar. Include negative controls and counterexamples by design.
  
  \item \textbf{Leakage and benchmark contamination checks:} Verify that no information from evaluation tasks, hidden test sets, or benchmark-specific artifacts can enter training, augmentation, hyperparameter tuning, or prompt design; when leakage is plausible, confirm results on stricter holdout or semi-private splits and disclose all benchmark-exposure pathways.
\end{itemize}

\textbf{Examples of Hypothesis Formation Advice:}
\begin{itemize}
  \item \textbf{Beware Bias for Elegant Algorithms:} Models may not prefer simple, elegant and human-understandable algorithms.
  \item \textbf{Models are lazy learners:} Models prioritise prediction accuracy over deep understanding. If there is a simple heuristic that allows accurate predictions, they will likely learn that and cease learning. As long as they get the right answer, they are done. 
  \item \textbf{A model algorithm sub-task tends to improve prediction accuracy:} A model only learns a subtask if it improves prediction accuracy - even if just a little in an edge case. This can give you clues on what sub-tasks might exist. e.g. in Addition, a sub-task to map pairs of digits (e.g. “2 with 4 maps to 6”, termed BaseAdd) will give the correct prediction if there is no carry-one from the previous column. So for some questions, it gives the correct answer. That makes it useful enough and the model learns it. 
\end{itemize}

\textbf{Examples of Auditing Red Flags:}

\begin{itemize}
  \item \textbf{Purely qualitative visuals or cherry-picked cases without falsification:} \citet{Adebayo2018} show that visually compelling explanations can be independent of model or data. Any claim that relies on visual appeal without randomization checks is insufficient.

  \item \textbf{Subjective human scoring as the primary validator:} Reviews emphasize that human evaluation is inconsistent and non-scalable. Use automated, preregistered tests and report uncertainty.


  \item \textbf{Uncalibrated localization:} If perturbations outside the circuit also change behavior at similar magnitudes, the claim lacks specificity (identify true negative circuits correctly). 


  \item \textbf{Benchmarks omitted or misinterpreted:} Ignoring community baselines leads to overclaimed novelty. 
\end{itemize}

\textbf{Examples of Experiment Goals:}
\begin{itemize}
  \item \textbf{Goal A:} Comparable claims across labs. Express every claim with the same primitives: a set of nodes, edges, interventions, and quantitative outcomes for preservation, localization, and minimality. Provide machine-readable artifacts so that automated verifiers can ingest results and recompute scores. 


  \item \textbf{Goal B:} Generalization beyond toy demos. Require that every confirmed claim passes the three property tests on out-of-distribution probes of the capability. 
\end{itemize}




\section{More Alternative Views}
\label{app:counters}

We continue addressing alternative views from Section \S \ref{sec:counters} in this section.

\noindent \textbf{View 3: Defining these guidelines will be difficult} 

A similar counterargument to \#2 is that there are tasks in mechanistic interpretability that are difficult to assess objectively and precisely. For instance, whether steering works well or not may be subject to arbitrary, case-specific thresholds that do not generalize.

Mirroring our previous response, we propose that these guidelines will be written not to define every assessment, but to ensure that the majority of mechanistic interpretability studies follow the best practices built on expert consensus that has been shaped by feedback from real-world applications. As such, they will avoid introducing unjustified rules, such as claiming that steering can only work well according to exact thresholds, if historical studies demonstrate that these thresholds depend on a specific case-by-case basis that are difficult and cumbersome to quantify. 

Instead, these guidelines will ensure minimal practices are followed, and if they are not or could not be (such as having barriers like being unable to apply a certain diagnostic method due to technical limitations), 
they will estimate the degree of uncertainty in hypothesized claims. 

For steering ``well", this means not forcing its definition to be within an arbitrary threshold, but to ensure that steering experiments are conducted on benchmarks that are representative of real-world scenarios for the target task (e.g., coding). Approaches do not are marked as ``largely uncertain".

\noindent \textbf{View 4: It is too early to introduce standards}

This argument states that the field is too young for standardization, and that premature formalization could restrict methods or discourage exploration. However, the intent of this proposal is not to impose a finished framework immediately. Instead, we call for a structured, collaborative process to begin developing one that starts with open discussion, evolving drafts, and iterative refinement.

It is precisely because the field is growing rapidly that initiating this effort early is advantageous. Early attempts will not be perfect, but they will surface areas of disagreement, reveal practical needs, and help shape a community record of arguments, revisions, and precedents. As AI capabilities advance, delaying the creation of verification infrastructure increases the risk that interpretability research will become too vast and heterogeneous to organize retroactively. 

\noindent \textbf{View 5: The field is evolving too fast for the existing techniques to be standardized} 

New techniques in MI are constantly being developed, and it can be argued that unlike in established practices such as medicine, these techniques will frequently be replaced such that developing standards for one technique will be useless as that technique will not be useful very quickly.

We propose that there are general approaches in MI which stand the test of time, and are unlikely to go away, such as activation patching, steering, and circuit discovery. Guidelines developed for these general approaches will also likely remain steadfast. As an analogy, many new software tools are constantly being developed, but general ``good coding and SWE practices" have endured.

\section{Importance of MI Auditing for Regulatory Policies}
\label{app:impacts}


Regulations impose outcome-oriented transparency requirements while remaining agnostic to implementation methods.
The EU AI Act ~\cite{eu_ai_act_2024} mandates transparency sufficient for deployers to interpret a system's output and use it appropriately (Article 13). However, accompanying guidance acknowledges the regulation does not clearly address the actual level of transparency required. Violations carry fines up to €35 million or 7\% of global annual turnover, whichever is larger.
~\citet{cfpb_circular_2023_03} articulates the most technically specific requirements, and rejects generic adverse action explanations, demanding that credit denials based on behavioral spending data identify which specific behaviors triggered the decision. ~\citet{cfpb_supervisory_highlights_2025} escalated the requirements: examiners now actively search for Less Discriminatory Alternatives using open-source tools, scrutinize models with over 1,000 variables for proxy discrimination, and require rigorous validation of adverse action methodologies.
The FDA's evolution illustrates regulatory maturation without mechanistic specificity. The most comprehensive FDA AI document to date ~\cite{fda_ai_lifecycle_guidance_2025} emphasizes clear interpretation mechanisms among its ten Good Machine Learning Practice principles ~\cite{fda_gmlp_principles} while acknowledging that authorized devices exhibit often limited explainability of AI predictions.

\section{MI Protocols}
\label{app:protocols}

A shared protocol format allows smoother application of guidelines to audit specific instances. For instance, if a guideline says "measure minimality of feature $X$ to test hypothesis $H$", knowing what specifically should be defined as feature $X$, and how to measure minimality, would be beneficial and reduce confusion.

For scalability and to mitigate human errors, experiments can include a protocol that defines how an automated AI system translates unstructured results into structured formats.
This protocol allows humans to work together with automated interpretability systems, to assess discoveries at a scale that manual efforts cannot. If each result is then expressed in the same schema, automated comparison and meta-analysis become feasible, turning a patchwork field into a coherent, cumulative discipline. We expect that such a protocol would not be immediately built, but will crystallize over time as guidelines become more established. An example of the layers of this protocol is given in Table \ref{tab:protocol}. 

\begin{table*}[h]
\centering
\begin{tabular}{|p{3cm}|p{4.5cm}|p{4.5cm}|}
\hline
\textbf{Protocol Layer} & \textbf{Purpose} & \textbf{Details} \\
\hline
Community Rules (Human Layer) &
Define norms, requirements, and minimum validity criteria &
A living ``standards document'' (e.g. web-wiki), community review processes, versioning, rule proposals \& voting \\
\hline
Abstract Framework / Schema &
Translate rules into a canonical formal language. Each author can define a schema &
Schema definitions (e.g. JSON), ``MI hypothesis'' primitives, required test definitions, metric templates \\
\hline
Machine-Readable Protocol &
Encode the framework so tools can parse, check, and verify claims &
APIs, validation scripts, automatic verifiers, benchmark suite \\
\hline
\end{tabular}
\caption{An example of abstraction layers in a MI protocol}
\label{tab:protocol}
\end{table*}

\section{Automated Auditing Assistance using Probabilistic Logic}
\label{app:autoaudit}

An automatic auditing system would parse the evidence provided for a hypothesized claim and suggest how certain that claim is, while identifying components that are incomplete or improperly executed. The protocol would format experimental data into a structured representation suitable for logical verification and iterative confidence estimation. 

Logical frameworks such as PSL can help filter out noise and curb the hallucinations that may arise in LLM-driven auditing \cite{logic_xu2023logiclm, logic_trinh2024solving}. We propose combining the two approaches: large language models would specialize in parsing and structuring unstructured experimental result data, while logical probabilistic frameworks such as PSL would act as verification layers that translate this parsed evidence into quantitative, internally consistent estimates of certainty. Together, this hybrid design would allow automatic auditing systems to interpret natural-language experiment descriptions and evaluate them through a rigorous, rule-based uncertainty model.

Recent progress in neurosymbolic reasoning provides a foundation for this approach. Systems such as \textit{Logic-LM}, which integrates symbolic solvers to improve the faithfulness of LLM reasoning \cite{logic_xu2023logiclm}, and \textit{LINC}, which couples first-order logic provers with language models for verifiable reasoning \cite{logic_2023linc}, demonstrate that hybrid reasoning pipelines can produce both interpretability and formal soundness. Parallel efforts in \textit{autoformalization} \cite{logic_weng2025autoformalization}, which aim to convert unstructured scientific or mathematical outputs into logical representations, further highlight the feasibility of this integration by bridging free-form LLM output with symbolic verifiers.

A particularly relevant precedent is the recent \textit{Nature} study on \textit{Solving Olympiad Geometry without Human Demonstrations} \cite{logic_trinh2024solving}, which showed that coupling LLMs with structured geometric solvers yields precise, verifiable reasoning in a complex symbolic domain. This work illustrates how a similar strategy could underpin automated auditing: by combining LLMs for linguistic and evidential parsing with probabilistic logic systems for formal evaluation, one can achieve both expressive understanding and rigorous verification of experimental claims.

\noindent \textbf{Example of using PSL to Assist with Automated Auditing. } 
In PSL, predicates (representing properties or relationships) take on soft truth values in [0,1], allowing degrees of truth rather than binary outcomes. Given experimental data, PSL infers predicate values and learns rule weights by optimizing a convex objective, aggregating disparate pieces of evidence into calibrated confidence.

We demonstrate an example with a sample set of predicates for a circuit minimality problem. Assuming we have some explicit performance metrics, we break down the problem into observed variables, and latent variables.

\noindent \textbf{Grade Mapping Predicates.}
We define a set of ``grade" predicates for the Minimality score: 
\begin{tabular}{ll}
\noindent \textbf{Predicate} & \textbf{Condition} \\
\hline
Min\_ge\_095(C) & $\text{Minimal}(C) \ge 0.95$ \\
Min\_ge\_090(C) & $\text{Minimal}(C) \ge 0.90$ \\
$\vdots$ & $\vdots$ \\
Min\_ge\_070(C) & $\text{Minimal}(C) \ge 0.70$ \\
Min\_ge\_060(C) & $\text{Minimal}(C) \ge 0.60$ \\
\end{tabular}

\begin{enumerate}
    \item {\textbf{InCircuit(E,C)}: Edge $E$ is in candidate circuit $C$ (binary variable).}
    \item {\textbf{Sufficient(C)}: Circuit $C$ meets the behavioral or specification threshold (binary).}
    \item {\textbf{Cost(C,V)}: Normalized cost or size of $C$ in $[0,1]$ (real-valued).}
    \item {\textbf{PerfDrop(E,C,V)}: Normalized performance drop in $[0,1]$ when removing $E$ from $C$ (higher $=$ worse performance).}
    \item {\textbf{RemMass(C)}: Precomputed or aggregated removable mass for $C$ in $[0,1]$ (e.g., the sum or average of $\text{Removable}(E,C)$ over all edges).}
\end{enumerate}

\noindent\textbf{Latent Targets.} These are the unobserved variables that must be inferred by the system:
\begin{enumerate}
    \item {\textbf{Critical(E,C)}: (\textit{latent}) Edge $E$ is necessary for the sufficiency of $C$ (degree in $[0,1]$).}
    \item {\textbf{Removable(E,C)}: (\textit{latent}) Removing $E$ preserves the sufficiency of $C$ (degree in $[0,1]$).}
    \item {\textbf{Minimal(C)}: (\textit{latent}) Circuit $C$ is approximately minimal with respect to cost, given sufficiency (degree in $[0,1]$).}
\end{enumerate}

\noindent \textbf{Grade Labels (latent)}

\begin{tabular}{l}
\textbf{Predicate} \\
\hline
Grade\_A(C) \\
Grade\_Aminus(C) \\
$\vdots$ \\
Grade\_D(C) \\
Grade\_E(C) \\
\end{tabular}

This provides a general recipe adaptable to other use cases. There are other alternative probabilistic and logic based systems, such as Dempster-Shafer Theory \citep{shafer1976mathematical}, that may apply to such systems, as described in Table \ref{tab:circuit_minimality_systems}. We encourage the community to develop and experimentally compare such variants across interpretability use cases, gradually identifying which frameworks best balance interpretability, scalability, and epistemic rigor in automated auditing.

\begin{table*}[ht]
\centering
\caption{Representative probabilistic and logic-based systems and how they might tackle circuit minimality auditing.}
\begin{tabular}{p{0.28\linewidth}|p{0.68\linewidth}}
\toprule
\textbf{Framework} & \textbf{Application to Circuit Minimality} \\
\midrule
\textbf{Probabilistic Soft Logic (PSL) \citep{psl_1}} & Represents sufficiency and edge importance as soft predicates inferred jointly, quantifying uncertainty about minimality under conflicting evidence. \\
\hline
\textbf{Markov Logic Networks (MLNs) \citep{MLN}} & Encodes weighted logical rules such as ``if edge $E$ is critical, removing it should reduce performance,'' estimating the probability that a candidate circuit is minimal given observed interventions. \\
\hline
\textbf{DeepProbLog \citep{NEURIPS2018_dc5d637e}} & Combines symbolic rules for minimality (e.g., necessity and sufficiency) with neural submodules that estimate probabilistic truth values from activation representations. \\
\hline
\textbf{Neural Theorem Provers (NTPs) \citep{rocktäschel2017endtoenddifferentiableproving}} & Differentiably matches logical patterns against learned embeddings of activation traces to infer relations such as ``edge $E$ contributes to function $F$.'' \\
\hline
\textbf{Causal Bayesian Networks (CBNs) \citep{gansch2025causalbayesiannetworksdatadriven}} & Represents directional dependencies between components and outputs, allowing interventions (e.g., ``remove $E$'') to estimate causal effects and test minimality. \\
\bottomrule
\end{tabular}
\label{tab:circuit_minimality_systems}
\end{table*}

\section{Auditing Minimal-Circuit Claims}
\label{app:minimal_llm}
Mechanistic interpretability papers frequently present explanations in the form of
\emph{minimal circuit stories}: a small set of heads, neurons, or features that are
claimed to implement a task. However, such explanations can be misleading if key
validity checks are missing. Inspired by recent discussions on auditing MI claims,
we propose a simple checklist that evaluates whether a circuit explanation satisfies
basic evidentiary standards.

To operationalize this idea, we implement a lightweight auditing tool that parses a
written explanation and checks whether it contains evidence for a set of minimal-circuit
guidelines. The tool prompts an LLM to evaluate whether the explanation provides
explicit evidence for each guideline.

\begin{tcolorbox}[colback=gray!5,colframe=black,title=Audit Prompt]
You are evaluating a mechanistic interpretability explanation that claims
(or might claim) to have found a minimal circuit for some task.

For each guideline below, determine whether the explanation provides clear
evidence that the guideline is satisfied. If evidence is missing or unclear,
mark it as false.

Guidelines include: sufficiency evidence, necessity evidence, sparsity of the
identified circuit, robustness on held-out data, causal validation through
interventions, sanity checks against null baselines, and stability across
random seeds or pruning tie-breaks.
\end{tcolorbox}

Figure~\ref{fig:scorecard-example} shows the output of the auditing tool on a toy
Indirect Object Identification (IOI) circuit explanation. The example satisfies
most minimal-circuit criteria but fails checks for multiple initializations and
tie exploration during greedy pruning.

\begin{table}[ht]
\centering
\renewcommand{\arraystretch}{1.2}
\begin{tabular}{p{5.5cm}c}
\textbf{Minimal-Circuit Guideline} & \textbf{Status} \\
\hline
Claims minimal/sparse circuit & \cmark \\
Sufficiency evidence & \cmark \\
Necessity evidence & \cmark \\
Sparsity / small circuit & \cmark \\
Robustness (held-out / shifts) & \cmark \\
Non-spurious (not cherry-picked) & \cmark \\
Causal validation (patching / ablations) & \cmark \\
Sanity checks / null baselines & \cmark \\
Multiple initializations / seeds & \xmark \\
Tie exploration in greedy pruning & \xmark \\
\end{tabular}
\caption{Example scorecard produced by the minimal-circuit auditing tool. The
explanation provides strong causal and robustness evidence but does not analyze
stability across seeds or tie-breaking choices in greedy pruning.}
\label{fig:scorecard-example}
\end{table}

This checklist is not meant to be exhaustive, but it provides a practical
framework for identifying common fallacies in circuit explanations. In
particular, missing seed robustness or pruning tie exploration can indicate
that the reported circuit may be unstable or one of several equally plausible
solutions.

\end{document}